\documentclass[a4paper,fleqn]{cas-sc}

\usepackage{lineno,hyperref}
\modulolinenumbers[5]

\usepackage[numbers]{natbib}

\usepackage{graphicx}
\usepackage{xcolor}

\usepackage{enumerate}

\newcommand{\beq}{\begin{equation}}
\newcommand{\eeq}{\end{equation}}
\newcommand{\bei}{\begin{itemize}}
\newcommand{\eei}{\end{itemize}}
\newcommand{\beqa}{\begin{eqnarray}}
\newcommand{\eeqa}{\end{eqnarray}}
\newcommand{\beco}{\begin{columns}}
\newcommand{\eeco}{\end{columns}}

\newcommand{\abs}[1]{\left\vert#1\right\vert}
\newcommand{\ket}[1]{\vert #1 \rangle}
\newcommand{\bra}[1]{\langle #1 \vert}
\newcommand{\mean}[1]{\langle #1\rangle}

\begin{document}


\shorttitle{Oscillators with \emph{Morse-like}  squared frequency $\omega^2(t)$ }
\shortauthors{M Gianfreda and G Landolfi}

\let\WriteBookmarks\relax
\def\floatpagepagefraction{1}
\def\textpagefraction{.001}

\title[mode = title]{Effects of quenching protocols based on parametric oscillators}

\author[1]{Mariagiovanna Gianfreda\fnref{mgfootnote}}[orcid=0000-0002-7352-2673]
\address{Department of Physics, Washington University, St. Louis, MO 63130, USA}
\fntext[mgfootnote]{Maria.Gianfreda@le.infn.it} 

\author[2]{Giulio Landolfi\fnref{glfootnote}}[orcid=0000-0001-6699-7876]
\address{Dipartimento di Matematica e Fisica "Ennio De Giorgi",  Universit\`a del Salento and INFN Sezione di Lecce, 73100 Lecce, Italy}
\fntext[glfootnote]{giulio.landolfi@unisalento.it, \sep giulio.landolfi@le.infn.it}

 \cortext[cor1]{Corresponding author: G. Landolfi}
 


\begin{abstract}
We consider the problem of understanding the basic 
features displayed by quantum systems described by parametric oscillators whose time-dependent frequency parameter 
$\omega(t)$ varies  continuously during evolution so to realise quenching protocols of different types. 
To this scope we focus on the case where $\omega(t)^2$ behaves like a Morse potential, up to possible sign reversion and translations in the $(t,\omega^2)$ plane. 
We derive closed form solution for the time-dependent amplitude of quasi-normal modes, which is the very fundamental dynamical object entering the description of both classical and quantum parametric oscillators, and highlight its significant characteristics for distinctive cases arising based on the driving specifics. After doing so, we provide an insight on the way quantum states evolve by paying attention on  the position-momentum Heisenberg uncertainty principle and the statistical aspects implied by  second-order correlation functions over number-type states.

\end{abstract}




\maketitle

 \section{Introduction } 
\label{s1}

		\setcounter{footnote}{0}

Time-dependent quadratic hamiltonians play a fundamental role in physics and have been therefore intensively studied   both at the classical and the quantum level.
As a matter of fact, they have widespread applications including quantum optics and quantum information  \cite{QOin, nerukh,pedrosa2011, kalluri, paul traps in,QOfin, hucul,  paul traps fin},  quantum cosmology \cite{sakharov, GRAVin, maggiore, geralico, trans, GRAVfin}, plasma physics \cite{pashaev, plasma in, plasma fin}, disordered systems \cite{disordered}, and Bose-Einstein condensates \cite{BECin, hu, yukalov, parentani,martone, BECfin}. Phenomena  thus appear in these areas of interest that share common features and similar originating mechanisms,  e.g. the  formation of wave patterns in spectra of dynamical variables and the creation of modes during the evolution of quantum fields. These aspects are currently inspiring experiments and the interplay between researchers involved in different areas. Indeed, the impressive progresses we are witnessing on the technological side open up to 
 the possibility to probe time-dependent quadratic hamiltonian models subjected to variant parametric drivings and  make more feasible the test of fundamentals of a theory,
 either directly or comparatively  \cite{plasma fin, dodonovdodonov, hung, ginis, nori3}. Interesting examples are provided by	 the electron-hole plasma created on the surface of a semiconductor slab by means of laser pulses,  an experimental setup that has been discussed as a mirror for the nonstationary Casimir effect inside a cavity with moving boundaries \cite{plasma fin, dodonovdodonov}, and by atomic superfluids whose interaction strength is modified by Feshbach tuning, such as in \cite{hung} where authors laser cooled and Bose condensed cesium atoms in an optical dipole trap and then generated a quench of the atomic interaction resembling the change of gravitational field after inflation.

The recent very much improved capability to control materials properties, and the consecutive growing general interest in implementing \textit{active} devices through 
time-varying media \cite{nerukh,  kalluri,zhang,lin} for  applications of various kind, motivate a systematized collection of cases study providing intelligible and accurate predictions 
in respect to the evaluation of the dynamical responses of parametric oscillator systems; see also the discussion within the Bose-Einstein condensation context \cite{martone}. 
This would enable also one to  gain a better knowledge of which conditions (initial state, control parameters) may turn out more advantageous for each definite application  (see e.g. discussions in the context of superconducting circuit technology \cite{nori3, nori1, nori2}) for which a quenching protocol is designed through parametric oscillators or related hamiltonian models as those in \cite{buyakasic}.  

In this paper we fix some specific time dependent frequencies $\omega(t)$ and address the study of one-degree of freedom quantum systems described by a parametric oscillator hamiltonian operator.  We are aimed at giving an account on the way on the oscillation mechanism that arise at the classical level for the amplitudes of non-stationary modes 
is influenced by deviation from adiabaticity,  how it is inherited at the quantum level, and the influence it has on quantum correlations.  Having this in mind, we shall perform therefore a study concerned with  another solvable example of a time-dependent quadratic problem: a
square parametric frequency functions $\omega^2$ assuming a form connected with that of Morse potential function.  Acting on a Morse potential
 shape through possible  time-translations and sign-reversions  will enable us 
to analyze, within  a  single analytical problem,  the role played by changes in the energy subtraction/pumping scheme, such as whenever parameter are chosen in a way that asymmetrical barrier or well show up and a monotonicity  is lost
for  $\omega(t)^2$, or giving rise to changes that perform non-adiabatically on a finite time-scale. 
We discuss extensively features of solution to the classical equation of motion, specifically their amplitudes $\sigma$. 
The step put in the position to build a solid  bridge leading to the quantum dynamics. 
While dealing with time-dependent quadratic Hamiltonian systems and   their dynamical comportments,  a convenient guideline 
is  the known fact that
their quantum dynamics is in actuality ruled by the classical one: expectation values follow the classical motion {\em exactly},  and the spreading of solutions to Schr\"odinger can be understood in classical terms; see e.g. \cite{littlejohn,andrews, dieter book,manko,olgamanko}. For this reason one should be well acquainted with 
features of the classical dynamics, and particularly  of amplitudes $\sigma$ of solution to the classical equation of motion. 
 Indeed, it is the amplitude function $\sigma$ that is later  recognized to play the pivotal role in the quantum  description of parametric oscillator dynamics.  
Functions $\sigma$ obey by the Ermakov differential equation 
\beq
\ddot{\sigma}(t)=-\omega(t)^2\, \sigma(t) +\frac{c}{\sigma(t)^3} \,\,  , \qquad \qquad  
\label{eq Ermakov}
\eeq
where dots denote time derivatives, $\omega(t)$ is the time-dependent driving frequency, and $c$ is a real  positive constant parameter (merely acting as a scale factor 
for $\sigma$). Solutions to Equation (\ref{eq Ermakov}),  and their time-differential $\dot{\sigma}$, specialize the classical system orbits in phase-space, and in turns this 
determines the elements  of the position-momentum correlation matrix among quantum wave-packets solving the Schr\"odinger equation \cite{dieter book, manko}. 
This is an instance of what is arguably the most profound connection of quantum mechanics with the classical one: the uncertainty principle, in its Robertson-Schr\"odinger 
strong form,  relies on the existence of Poincar\'{e} invariants on the classical side \cite{robertson, narcowich, manko, de gosson, eliashberg}, the Planck's constant simply 
representing a scale requirement.   For the one degree of freedom systems ruled by the time-dependent parametric oscillator hamiltonian, the basic invariant is the area
 the ellipses in phase space determined for assigned values of the so-called Ermakov invariant,  a quadratic N\"other  invariant for the system \cite{profilopra}. 
Their deformation in the course of dynamics is ruled by the amplitude $\sigma$ changes, whose oscillatory features intertwine with those of classical Newton  equation of 
motion for the parametric oscillator, a Sturm-Liouville problem indeed. Oscillations for classical orbits consequently drive the squeezing on cognate Wigner ellipses at 
the quantum level.  Peculiar oscillating phenomena therefore enter quantum correlation functions and mode-counting statistics. Examples will be given 
of highly oscillating amplitudes determining strong squeezing phenomena in phase-space, and of the manner they  steer the mode-counting statistics. 

The outline of the paper is as follows. In Section 2, we provide a recap of the treatment of time-dependent quadratic  quantum systems through the Lewis-Riesenfeld invariant approach \cite{c2}.  After recalling the linkage between the parametric oscillator and the Ermakov equations, in Section \ref{ss 3} we make some general comments on the features
of their solutions. In Section \ref{ss 5},  we present figures showing solution to Ermakov equation based on suitable initial conditions and representative sets of  parameters
entering the time-frequency function $\omega(t)$ we opted for, and analyze thoroughly the features displayed. In Section \ref{ss 6} we pay attention on implied squeezing 
of classical  Ermakov ellipses in phase-space,  which really means to comprehend  the Heisenberg uncertainties  for position-momentum operators
 at the quantum level. In Sections \ref{ss 7} and \ref{ss 8} our attention is devoted to mode-counting statistics. There, 
 we consider normalized second order correlation functions pertinent nonclassical states of the number type and argue on the typicalities that are visible from their plots.   
Section \ref{ss 9} will be for conclusions. Finally, in Appendix \ref{appendice},  we  expound the derivation of the solutions to the  parametric oscillator equation for the frequency parameter we have chosen and determine their Wronskian, a  task  needed to the construction of analytical  solutions to 
 Ermakov equation (\ref{eq Ermakov}).

\section{ Quantum parametric oscillator: a brief on the theoretical background}
\label{s2}

The quantization of time-dependent quadratic systems is a well established topic, see e.g. \cite{manko, dieter book}. 
The most effective way to proceed has been recognised in the approach proposed by Lewis and Riesenfeld in \cite{c2},  
a method  intrinsically relying on the identification of symmetries and invariants.
In this Section we recall the basic steps connecting the solutions to the Schr\"odinger equation pertinent to a parametric oscillator hamiltonian operator to the spectral problem for the quantum Ermakov invariant operator. After doing so, the Bogolubov map relating the invariant description to Sch\"rodinger picture is presented. This will enable us to  delineate the squeezing dynamics experienced by the system.

\subsection{The Lewis-Riesenfeld approach to quantum 
parametric oscillator}
\label{ss LR}

 Consider the time-dependent one-dimensional Hamiltonian 
\beq
H(t)=\frac{p^2}{2\,m}\,+\,\frac{m\,\omega^2(t)}{2}q^2,
\label{eq2}
\eeq
where $q$ and $p$ are operators that satisfy the canonical commutation relation $[q,p]=i\hbar$. The mass parameter $m$ is not time dependent. 
For systems governed by the hamiltonian operator (\ref{eq2}), a convenient description can be given in terms of a time dependent hermitian operator $I \equiv I(t)$ which is an invariant; that is, $ I^\dagger=I$ and $\dot{I}=\partial_t  I +(i \hbar)^{-1}\,[I,H]=0$, where the dot stands for the total time derivative. 
A state vector $|\psi(t)\rangle$  that satisfies the Schrodinger equation for (\ref{eq2}),
$i \, \hbar\, \partial_t  |\psi(t)\rangle=H(t)|\psi(t)\rangle$, can be written in terms of the time-dependent eigenstates $|\phi_n(t)\rangle$  associated to the invariant $I(t)$ with time-independent eigenvalues $\lambda_{n}^{(I)}$, $I\, |\phi_n(t)\rangle=\lambda_{n}^{(I)}|\phi_n(t)\rangle$,
via the general superposition formula $|\psi(t)\rangle =\sum_{n=0}^\infty c_n\,e^{i\,\alpha_n(t)}|\phi_n(t)\rangle$,
where the phase  $\alpha_n(t)$ satisfies the equation
\beq
\hbar\,\dot{\alpha}_n \,=\,\mean{ \phi_n \abs{\,i\,\hbar \,\partial_t-H} \phi_n }.
\label{eq7}
\eeq
By assuming that  $I(t)$ can be written as a quadratic expression in the dynamical variables $q$ and $p$, one can find that 
\beq
I=  \frac{ c}{\sigma^2(t)}q^2+\left[\sigma(t)\,\frac{p}{m}\,-\,\dot{\sigma}(t)\,q\right]^2 ,
\label{eq8}
\eeq
where $\sigma(t)$ is a real function of time, solution to the Ermakov  differential equation (\ref{eq Ermakov}).
Operator (\ref{eq8}) is essentially the quantised version of the classical invariant discussed originally by Ermakov in his prime analysis in \cite{ermakov}. 
The invariant operator $I(t)$ in Equation (\ref{eq8}) can be written as 
\beq
I(t)= \frac{2\,\sqrt{c}\,\hbar}{m}\left[a^\dagger(t)\,a(t)\,+\frac{1}{2}\right]
\,\, , \qquad \quad
a(t)=
\frac{1}{\sqrt{2\hbar}}\left[\frac{c^{1/4}\sqrt{m}}{\sigma}\,q\,+\,i\,\left(\frac{\sigma}{\sqrt{m}\,c^{1/4}}\,p\,-\,\frac{\sqrt{m}\,\dot{\sigma}}{c^{1/4}}\,q\right) \right]
\label{eq10}
\,\, .
\eeq
Operators $a(t)$ and $a^{\dagger}(t)$ are time-dependent operators of the annihilation and creation type, satisfying the canonical commutation rule $[a(t),a^\dagger(t)]=1$.  
Accordingly, $N=a^\dagger(t)\,a(t)$ is an  operator of the number type, 
from which it follows that the spectrum of $I(t)$ consists in the set of stationary eigenvalues $\lambda_{n}^{(I)}= 
\sqrt{c} \,m \, \hbar \, (n+\frac{1}{2})$, for $n=0,1,2\dots$, and the eigenstates are the same of the number operator $N$, $|\phi_n(t)\rangle \equiv\ket{n}$ being $N \ket{n}=n\ket{n}$.
Otherwise said, the  invariant operator $I(t)$ can be mapped into a time-independent harmonic oscillator via a unitary transformation (see e.g. \cite{mostafazadeh}).
This implicitly states that  Fock-type spaces
are defined at different times in terms of the common eigenstates $\ket{n}$ of the number operator $N$ and Ermakov invariant $I$, and that these spaces can be then mapped one into the other through a time-dependent transformation. 
 Similarly, coherent-like bases at different times may be defined by means of the eigenstates $\ket{\alpha} $ associated with the spectral problem for the time-dependent operator $\hat a$, \textit{i.e.} $\hat{a}\left| \alpha\right\rangle=\alpha\left| \alpha\right\rangle$. These are the so-called Lewis-Riesenfeld coherent states (thought they are actually squeezed states), and are expressible in terms of the states $\ket{n}$ through the usual relationship.
 The number-type states $\ket{n}$, and consequently the coherent-type states {\em a l\'a} Lewis-Riesenfeld $\ket{\alpha}$, do not solve the Schrodinger equation for the Hamiltonian operator   (\ref{eq2}). For this to happen the action of a unitary operator is needed that merely equips each component $\ket{n}$  with the proper time-dependent  phase $\alpha_n$
 defined via (\ref{eq7}),  an operation that do not affect means over  $\ket{n}$ or $\ket{\alpha}$
 and let  these states be of interest for practical purposes. 
 As in the standard harmonic oscillator case, exact analytical solutions in the form of gaussian wave-packets can be therefore found for the Schr\"odinger equation associated to a parametric oscillator hamiltonian. 
A strong connection is consequently established between classical and quantum dynamics. In fact,   the time-evolution of the maximum
and the width of a wave-packet  (the two quantities that uniquely determine it) are governed by classical dynamics: the maximum  behaves like a particle following just the classical trajectory and the width is proportional to the amplitude $\sigma$ of solution to classical Newton-Lagrange equation for the parametric oscillator \cite{ littlejohn,dieter book}.

\subsection{Unitary Bogolubov transformation for annihilation-creation operators}
\label{ss bogolubov}

The identification of features associated with the dynamics for the quantum  systems described by the hamiltonian (\ref{eq2}) can be addressed by resorting to a squeezing formalism \cite{walls}. This can be done by writing 
\beq
a(t)=\mu(t)\,a_0\,+\nu(t)\,a_0^\dagger, \qquad \quad |\mu|^2-|\nu|^2=1
\label{bogolubov  mapping a}
\eeq
where $a_0$ and its hermitian conjugate  $a^{\dagger}_0$ are the ladder-type operators applicable to the fixed-time hamiltonian operator  at an initial instant  $t_0=0$, i.e.
  \begin{align}
& 
H_0\equiv H(0) =\frac{p^2}{2 m}+\frac{1}{2}\,m\,\omega_0^2\,q^2=\hbar \, \omega_0\,\left(a^\dagger_0 a_0\,+\,\frac{1}{2}\right)
\,\, , 
\qquad \quad
\omega_0\equiv\omega(0) \,\, , 
\label{eq19} \\
&
a_0=
 \sqrt{ \frac{m\,\omega_0}{2\hbar}}\,q\,+\,\frac{i}{\sqrt{2\hbar m\,\omega_0}} \,p\,\,  .
\label{eq18}
\end{align}
The constraint $|\mu|^2-|\nu|^2=1$ on Bogolubov coefficients $\mu,\nu$ follows from the unitarity of transformation. We have:
\beqa
\mu(t)=
\frac{1}{2\,c^{1/4} \sqrt{\omega_0}} \,\left(  \frac{\sqrt{c}}{\sigma} +\omega_0\,  \sigma -i \dot{\sigma }\right) \,\, , 
\qquad \nu(t)=
\frac{1}{2\,c^{1/4} \sqrt{\omega_0}} \,\left(  \frac{\sqrt{c}}{\sigma} -\omega_0\,  \sigma -i \dot{\sigma }\right) \,\, 
\label{bogolubov mapping mu nu}
\eeqa 
(close formulae can be given when the mass-type parameter is time-dependent  \cite{la ru so};  adaptation to SUSY-QM formalism can be found in \cite{gianfreda tmp}).

Any element $\ket{m}$ of the  Fock space   determined by the invariant $I(t)$ spectral problem at time $t$, 
\textit{ i.e.}  a state $\ket{m}$ obtained by application of $(\hat{a}^\dag)^m$ on the $\hat{a}$-vacuum $\ket{0}$, 
can be expressed as an infinite superposition of the number states 
$\ket{n}_0$ belonging to the Fock space  
associated with the operators $\hat{a}_0$ and $\hat{a}_0^\dag$, for which $\hat{N}_0 \ket{n}_0=\hat{a}_0^\dag  \hat{a}_0 \ket{n}_0=n \ket{n}_0$.  The fundamental   overlaps $_0 \mean{ n | m }$ are established via 
Eqs. (\ref{bogolubov  mapping a}) and (\ref{bogolubov mapping mu nu}), and so the evolution of arbitrary 
wave-functions can be obtained accordingly in either of the basis 
(for the explicit form of states  $\ket{m}$ in position representation, see e.g. \cite{dieter book, manko,  pedrosa2011}).
Notice that at this stage there is a freedom in respect to the choice of initial condition for the Ermakov differential problem (\ref{eq Ermakov}), and this also affect the explicit form of Bogolubov mapping.  Unitarily inequivalent time-dependent  vacua  can be thus introduced via $a(t) \ket{0}=0$  by the  general form of the invariant factorization, 
Eq (\ref{eq10}). This is not surprising, as it generalizes to the non-autonomous case the well understood  unitary action of squeezing of orbits for the stationary case. 
To remove the vacuum ambiguity, in this paper we shall proceed in the most economical way, i.e. by demanding that  at initial time $t=0$ the operator $a(0)$  coincides with $a_0$ in (\ref{eq18}):  $\mu(0)=1$ and $\nu(0)=0$.
 This requirement   fixes initial conditions to the Ermakov  equation (\ref{eq Ermakov}) as follows:
\beq
\sigma_0\equiv \sigma(0)=\frac{c^{1/4}}{\sqrt{\omega_0}},\qquad\ \dot{\sigma}_0\equiv \dot{\sigma}(0)=0.
\label{eq21}
\eeq
Evidently,  the Hamiltonian and the invariant operators do not generally commute, $[I,H]=(\mu \nu^* \hat{a}_0^2-h.c.)$. But once initial conditions (\ref{eq21}) are superimposed 
Lewis-Riesenfeld states at initial time coincide with standard states for the  fixed-time harmonic oscillator with frequency $\omega_0$. 
Of course other choices can be made for $\mu(0)$ and $\nu(0)$ that more conveniently may allow to follow the evolution of  other states.

\section{Solutions to Ermakov equation: general aspects}
\label{ss 3}

Ermakov systems have  widespread occurrence mathematics and physics \cite{dosly,  athorne,  rogers, rogers schief,   leach, matzkin, thylwe, auzzi} 
as they naturally arise once,  in the spirit of the pivotal Ermakov and B\"ohl discussions \cite{ermakov, bohl},  solutions to second order linear differential equations are 
expressed via the sine and the cosine functions \cite{goff}.  In practice, the  Ermakov equation (\ref{eq Ermakov})  is nothing but one of the two differential constraints 
into which the  classical parametric oscillator equation 
\beq
\ddot{x}(t)\,+\,\omega^2(t)\,x(t)=0 \, \, 
\label{eq27}
\eeq
splits once solutions are sought in the  quasi-normal mode form 
 \beq
 x=  \frac{\sqrt{I_{cl}}}{\sqrt{c}} \, \sigma(t) \cos[\theta(t)+\overline{\theta}_0]
\label{x sigma theta}
 \eeq
with real constants $\overline{\theta}_0$ and $I_{cl},\,c>0$, 
the other differential constraint to be satisfied for consistency determining the phase function via $ \dot{\theta}=\sqrt{c} \, \sigma^{-2}$. 
The constant $I_{cl}$ in (\ref{x sigma theta}) corresponds to the value assumed by the Noether invariant of the theory  obtained by taking $q$ and $p$   in (\ref{eq8})  as 
the classical position and momentum phase-space coordinate. Relying on this connection between the two differential problems, it is possible to employ solutions to 
the linear parametric equation to the purpose of desuming solutions to the nonlinear Ermakov amplitude equation   (\ref{eq Ermakov}). 
In exact terms,  the general solution to  Eq. (\ref{eq Ermakov}) can be expressed  in terms of two linearly independent solutions $x_1$ and $x_2$ of  (\ref{eq27}) via 
\beq
\sigma=\sqrt{A\,x_1^2\,+\,2\,B\, x_1\,x_2\,+\,C\,x_2^2}\,\,,
\label{eq30}
\eeq
with the condition 
\beq
A\,C - B^2=\frac{c}{W^2(x_1,x_2)}  \geq 0 \,\, ,
\label{eq31}
\eeq
where $W$ is the Wronskian of $x_1$ and $x_2$.  The  Ermakov-B\"ohl-type  formulae (\ref{eq30})-(\ref{eq31}) generalize those with $B=0$  considered at early stages \cite{bohl,ermakov},  and can also  be derived brightly using projective geometry \cite{athorne}.  
The constants $A,B,C$ can be completely fixed by imposing (\ref{eq31}) along with initial conditions  on $\sigma$ and $\dot{\sigma}$. 
Real solutions of Ermakov equation are thus positively defined and  do not vanish. 

The behavior of solutions to the Ermakov equation (\ref{eq Ermakov}) is obviously a reflex of features that are typical of solutions to the parametric oscillator
differential equation (\ref{eq27}). The presence of oscillating patterns for $\sigma$ can be awaited on the basis of the fundamental results obtained 
for equations of the parametric oscillator type  \cite{kong, zettl, bellman}. 
The well-known Fite-Leighton criterion, for instance,  
states that if $f(t)$ is a nonnegative real function  in the interval $t\in[t_0,\infty)$ such that  
${\rm lim}_{T\to\infty} \int_{t_0}^T f(t) \, dt=\infty$ then the equation $ \ddot{x}(t)+f(t)\,x(t)=0$ is {\em oscillatory}, 
i.e. has arbitrarily large zeroes on the independent variable domain $[t_0,\infty)$. 
When $f(t)$ is allowed to assume negative values for arbitrarily large values of $t$, the 
Wintner theorem can be applied, according to which if ${\rm lim}_{T\to\infty} \, (1/T) \,  \int^T_{t_0} dt \int_{t_0}^{t} f(t') dt' =\infty$
the equation is oscillatory. Furthermore, the Sturm comparison theorem
supports the conclusion that the larger is the parametric frequency the more rapidly  solutions to the parametric oscillator equation will oscillate, 
and the Sturm separation theorem  guarantees that given two linear independent solutions of the parametric oscillator equation the zeros of the two solutions are alternating.  
These results provide a convenient platform for appraising implications for the solutions to the Ermakov equation 
owing to the connection established via (\ref{eq30})-(\ref{eq31}).
Independent solutions for the parametric oscillator equation that alternate zeros introduces indeed a mechanism where terms in (\ref{eq30}) alternate minor and major contribution, and 
the function $\sigma$ consequently displays undulations that can be sustained to different  degrees depending  the specific form of independent solutions (i.e. on the shape of 
$\omega(t)$) and on coefficients $A,B,C$ (i.e. on the initial conditions for the Ermakov equation; see also the discussion in  \cite{cruz} appertaining the constant frequency case). 
In respect it is worth to recall that the inequality  
 \beq
 |x|\le (\, |x(t_0)|+|\dot{x}(t_0)|\, ) \,\,  e^{   \int_{t_0}^t  |f(s)-1|\, ds }\,\, .  \label{boundness inequality} 
 \eeq
(in proper units) generally holds \cite{bellman}. The polar form  of Ermakov-B\"ohl  decomposition   (\ref{x sigma theta})  then allows to transfer the inequality (\ref{boundness inequality}) directly to an inequality for the amplitude $\sigma$ comprising initial conditions for the Ermakov equation.  
Adiabaticity of variations for the frequency parameter $\omega$ do not suffice to guarantee that even changes for $\sigma$  are realised slowly. 
Attention has to be paid on the right-hand side of Equation (\ref{eq Ermakov}).
Demanding its vanishing  functions as retaining the leading order in the WKBJ expansion treatment of the parametric equation \cite{bellman, bender}.  
As long as $ \omega^2 \sigma^4$ deviates from attaining values about $c$,  the dynamical changes in amplitude $\sigma $  are  less smooth. 
 Accordingly, phases $\theta$ deviate from the nearly linear behavior of the adiabatic regime (bear in mind that $\theta$ directly determines the quantum phases resulting 
 from (\ref{eq7}) over number-type states through to the relation $\dot{\alpha}_n(t)=-(n+1/2) \sqrt{c} \, \sigma^{-2}$ \cite{c2}, i.e. $\dot{\alpha}_n(t)= -(n+1/2)\, \dot{\theta}(t)$).

\section{Analysis of amplitudes $\sigma$ for oscillators with Morse-type square frequencies}
\label{ss 5}

\begin{figure}[h!]
\begin{center}
\includegraphics[height=3.2 cm]{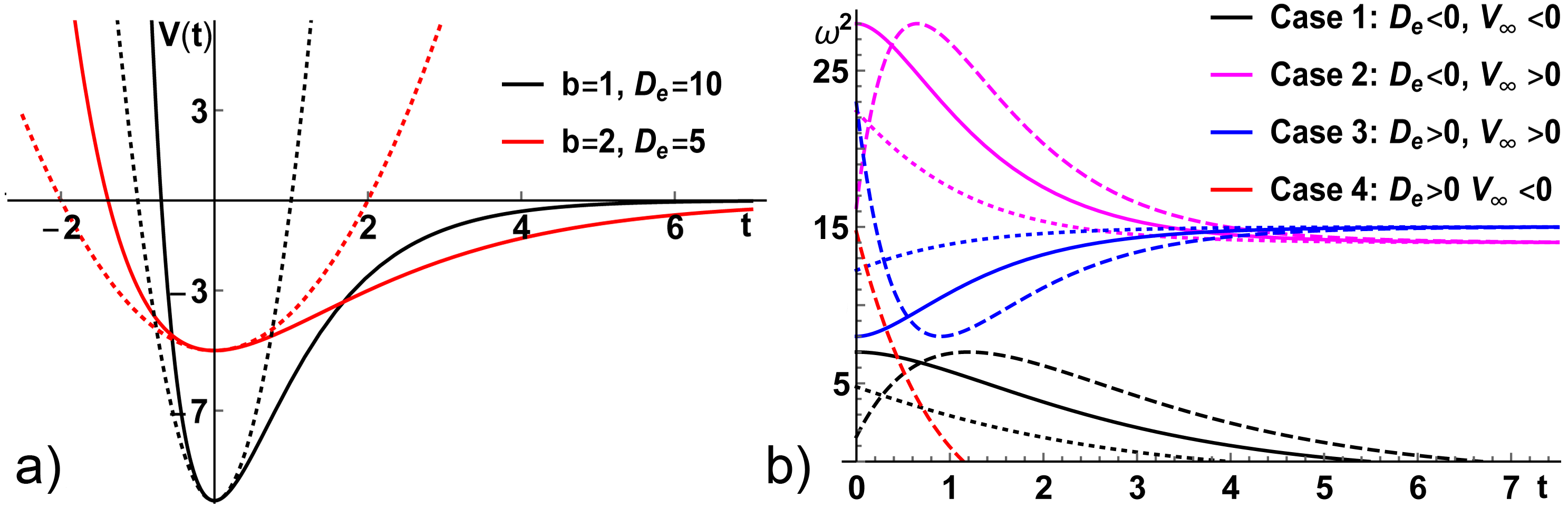}
\end{center}
\caption{
a) Morse potential  $V(t)=D_e \, (e^{-\frac{2\,t}{b}}\,-\,2\, e^{-\frac{t}{b}} ) $ compared to its harmonic approximation (dotted).
b) Examples of admissible $\omega^2(t)$ once all the possible actions on the Morse potential
 (sign reversion, vertical and horizontal shifts) are put forward  via Eq. (\ref{omega2 gen}); 
solid,  dashed and dotted curves refer to null, negative and positive time-translation parameters $t_s$. }
\label{figura morse}
\end{figure}

In this communication we consider a time-dependent quadratic system (\ref{eq2}) 
with square parametric frequency 
\beq
\omega^2(t)= D_e\,\, \left[e^{-\frac{ 2\,(t+t_s)}{b}}\,-\,2\, e^{-\frac{t+t_s}{b}}\right] +V_\infty \,\, , 
\label{omega2 gen}
\eeq
where $D_e\neq0$, $b>0$ and $V_\infty\neq0$ are real parameters. The form of Equation (\ref{omega2 gen}) basically represents a generalization in the time-domain
of  the well-known  Morse potential, which is indeed recovered  for $V_\infty=0$, where  the  parameters $D_e$ and $b$, both positive, determine the depth and the effective width of the Morse potential hole, see Figure \ref{figura morse}.
Adopting (\ref{omega2 gen}) we also consider a possible  upside-down flipping caused by the change of sign of $D_e$, as well as  horizontal translation of the standard Morse curves  via $t\to t+t_s$.  This enables us to  design different types of drivings, and to argue on the problem with a wider generality.  
 To this, it is convenient (especially for the analytical treatment, Appendix \ref{appendice}) to distinguish four cases
 relying on the signs of parameters $D_e$ and $V_\infty$:  
\textit{i)} Case 1, $D_e\,,\,V_\infty<0$;  \textit{ii)} Case 2, $D_e<0$, $V_\infty>0$;  \textit{iii)} Case 3, $D_e\,,\,V_\infty>0$;   \textit{iv)} Case 4, $D_e>0$, $V_\infty<0$.  
As we will see, further restrictions on parameters can arise because 
we shall be obviously concerned only with the cases for which   $\omega(t)\geq0$, and it shall be understood that   
 \beq \omega_0\equiv\omega(0)=\sqrt{D_e (e^{-2t_s/b}  -2e^{-t_s/b}  )+V_\infty} \eeq
   is strictly positive. The  qualitative behavior  for the $\omega^2$'s we shall deal with separately in each case is outlined in Figure \ref{figura morse}.b).\footnote{Actually, curves like the red one in Fig. \ref{figura morse}.b)    can be obtained also in Case 1  for extremely large negative $D_e$. 
  However, for the sake of an advantageous schematic  that  makes a more net contradistinction between  drivings with different features,  we shall exclude these kind of choices for   Case 1 and shall designate Case 4 for the analysis of dynamics under strongly non-adiabatic decreasing drivings ceasing in finite time. }

  When $V_\infty <0$  the dynamical study can be carried out  thoroughly for all times $t\geq0$ (Cases 2 and 3), otherwise   a positive  time-dependent square frequency rules the dynamics only over a finite time window (Cases 1 and 4) 
whose duration is determined through the root of equation $\omega^2=0$ defined via
\beq
t_+ =-t_s - b \log \frac{ D_e + \sqrt{ D_e (D_e-V_\infty) } }{D_e} \,\, .
\label{root tm}
\eeq

 In this Section, we analyse solutions $\sigma $ to Ermakov equation (\ref{eq Ermakov})  when $\omega^2$ takes the Morse-like form (\ref{omega2 gen}) 
and the parameters involved  are varied to modify width and height of barriers/holes as well as the position of the  curve 
in the $(t,\omega^2)$ plane. Attention will be also paid on phases $\theta=\sqrt{c} \int^t_0 \sigma^{-2} dt$. 
The examination can be conducted being conscious that the Fite-Leighton criterion  applies  in a natural manner to Cases 2 and 3,   for which $V_\infty,\, b >0$, and 
it is legitimate  to consider the asymptotic limit $t\to \infty$, that yields indeed to  ${\rm lim}_{T\to\infty} \int_{t_0}^T \omega^2(t')\, dt'=\infty$. 
In Cases 1 and 4, application of Wintner theorem interdicts the independent solutions to the 
equation $\ddot{x}+\omega^2x=0$  from possessing arbitrarily large zeroes. Nevertheless, in our study this differential problem is actually posed in a finite interval for the time independent variable. Zeroes in this interval are not excluded and,  depending on the manner initial conditions and parameters are fixed, 
visible fluctuations  may be produced. Also notice that the outcome of integration in  (\ref{boundness inequality}) (with parametric frequency cast in adimensional units) is always finite, even as $t\to \infty$ (in Cases 2 and 3 where the limit makes sense).  Hence solutions  to parametric and Ermakov equations are bounded,  but initial conditions may determine high upper bounds even when small variations of $\omega$ take place.

Features of the solutions to Ermakov equation in each of the four cases will be argued in the following Subsections
with plots summarising their behavior for different values of the parameters involved and the initial conditions (\ref{eq21}).
 The  implication at the classical level of such "instantaneous no-squeezing" condition is that it shapes parametric oscillator solutions (\ref{x sigma theta}) 
that at the initial time reproduce what one gets at the same time for the standard representation of stationary modes for  harmonic oscillators with mass $m$, proper frequency $\omega_0\equiv {\omega(0})$, constant amplitude and phase depending linearly on time. In different words, the curve determined in phase-space by the Ermakov invariant  at $t=0$ is just a rescaling by a factor $2\sqrt{c}/m\omega_0$ of the elliptic orbit of the harmonic oscillator with frequency $\omega_0$ and mass $m$, of which it preserves the minor/major axes ratio.

The following forewarnings are in order. 
Firstly, to avoid that the more mathematical details break the overall flow and affect the readability of the paper, the derivation of the formulae that are
necessary to construct   closed form solutions to the Ermakov equation (\ref{eq Ermakov}) via Eqs. (\ref{eq30})-(\ref{eq31}) will be 
worked out separately in Appendix \ref{appendice}. 
\footnote{Assuming the initial conditions (\ref{eq21}), for the constants in Equation (\ref{eq30}) one has: 
$\omega_0 \, \, W(x_1,x_2)^2\,\, A=    \sqrt{c}\,\, [\omega^2_0\, x_2^2(0)\,+ \dot{x}_2(0) ^2]$, 
$\omega_0 \, \, W(x_1,x_2)^2\,\,B=   \sqrt{c}\,\, [\omega^2_0\, x_1(0)\,x_2(0)\,+ \dot{x}_1(0)\, \dot{x}_2(0)]$, and  
$\omega_0 \, \, W(x_1,x_2)^2 \,\, C= \sqrt{c}\, \,  [\omega^2_0\, x_2^2(0)\,+ \dot{x}_2(0) ^2]  $. 
}
There the independent solutions to Equation (\ref{eq27}) 
are shown to be given in terms of the Tricomi confluent hypergeometric function $U$ and the Kummer function of the first kind $_1F_1$, whose Wronskians are next computed. 
Secondly, being beneficial to the physical interpretation of results, hereinafter a reference to the frequency $\omega(t)$
will often be present in plots through of dotted curves of the color of the principal quantities that will be evaluated with the same assignment of parameters; 
the way the reference to $\omega(t)$ will be given is established case by case depending on individual plot's convenience. 
Finally,  from now on all parameters entering the definition of the parametric frequency will be considered in adimensional units   
and it will be set $c=1/4$.

\subsection{Case 1: $D_e, \, V_\infty<0$}
\label{ss 5 1}

 The first case that we consider is a square parametric frequency defined through an inverse Morse potential with a negative nonvanishing asymptotic value. 
It is realized with negative parameters $D_e$ and   $V_\infty$, and such  that curve (\ref{omega2 gen}) intercepts the positive time axis once. 
   The full set of parameters involved in  (\ref{omega2 gen}) has also 
to guarantee that $\omega_0^2>0$ (i.e. a correctly defined cognate harmonic oscillator at the initial time). Thence the two roots  of $\omega^2=0$ have to be real, noncoincident   
and opposite in sign. That means   $D_e < V_\infty<0$ and $- b \log (1+\sqrt{1-V_{\infty}/D_e})<t_s<- b \log (1-\sqrt{1-V_{\infty}/D_e})$.
\begin{figure}[h!]
\begin{center}
\includegraphics[height=3.8 cm]{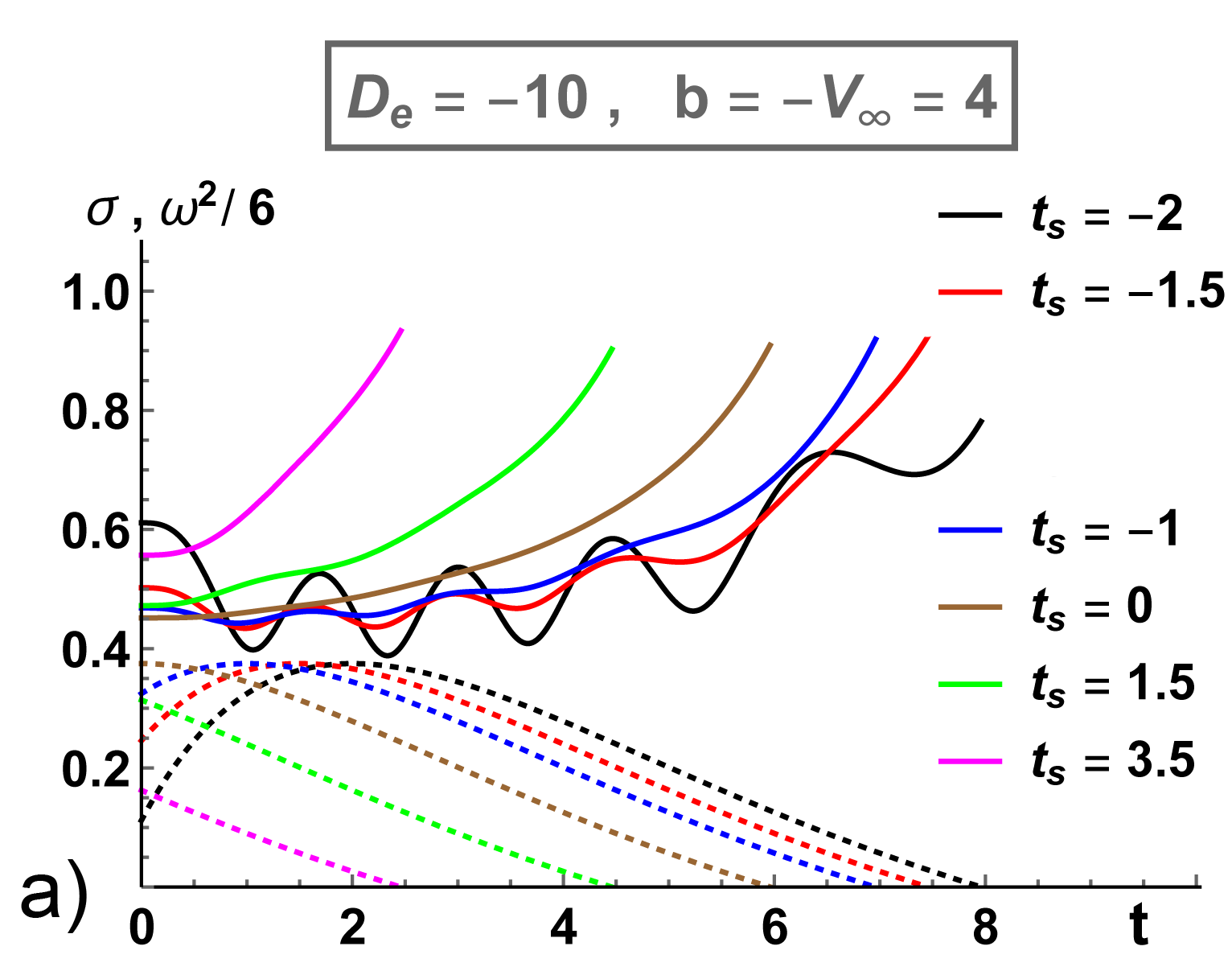} \quad
\includegraphics[height=3.8 cm]{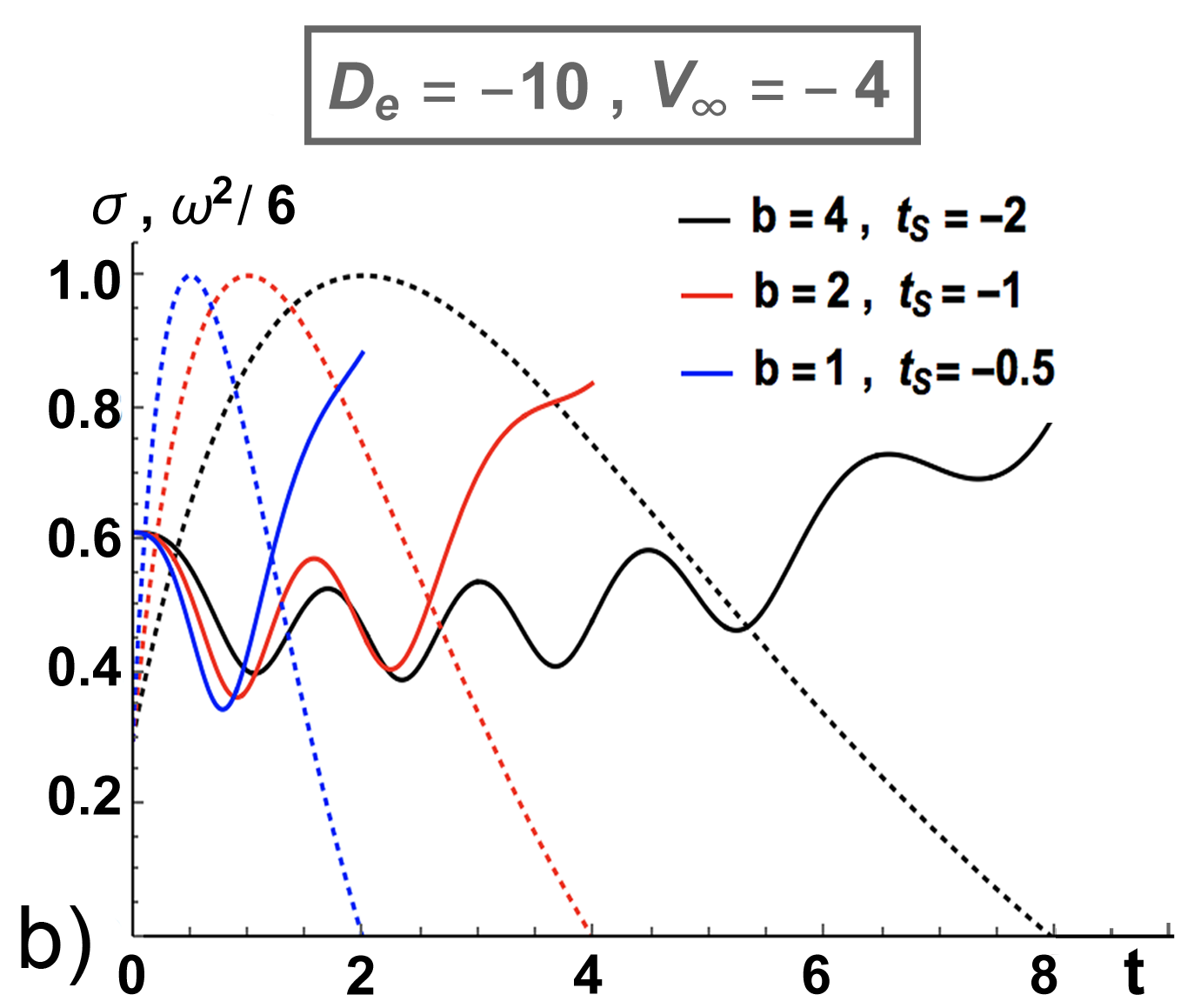} \quad
\includegraphics[height=3.8 cm]{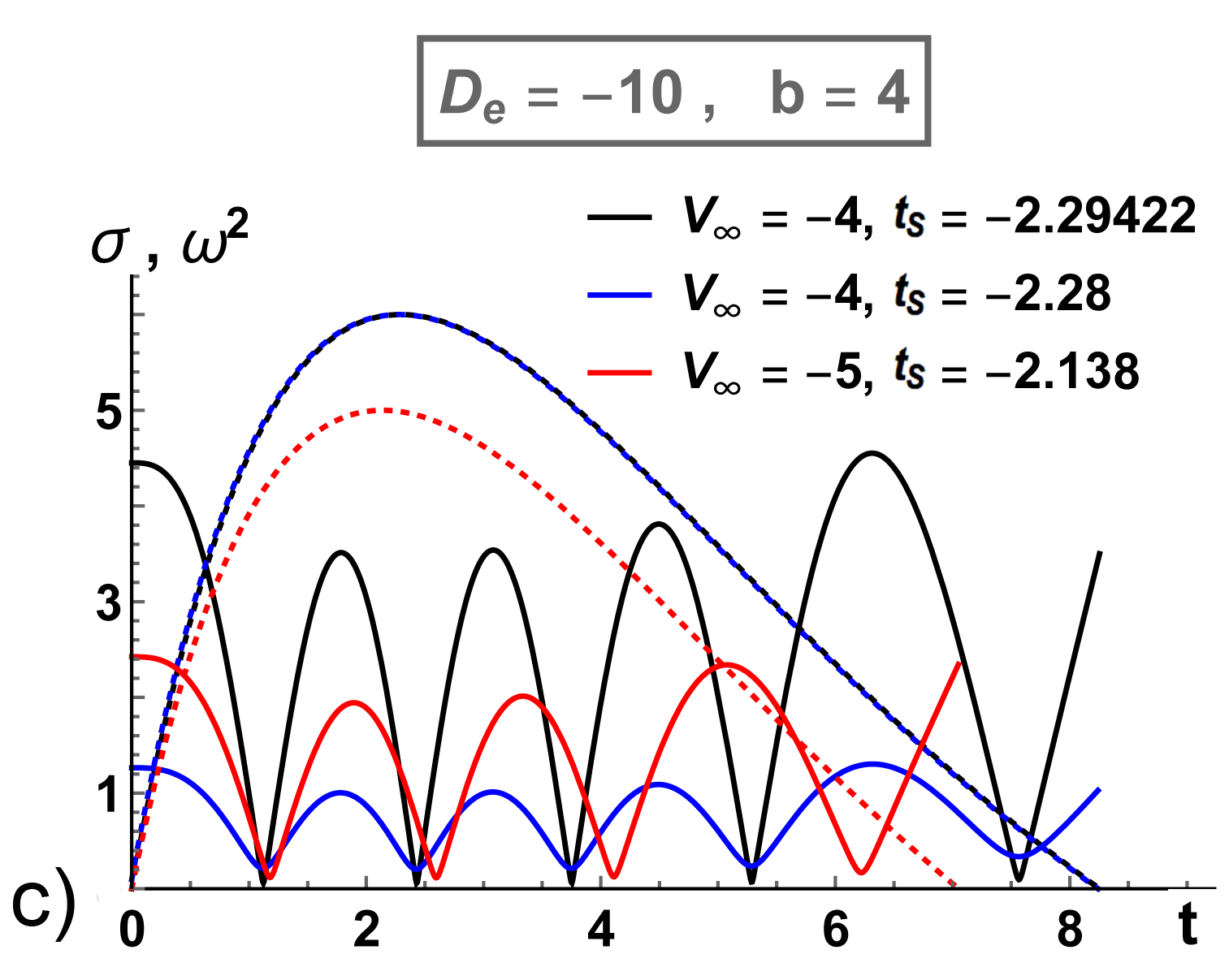} \quad
\end{center}
\caption{
Function $\sigma$ when $\omega(t)>0$ is defined via Eq. (\ref{omega2 gen}) with $D_e,V_\infty>0$. 
Dotted curves refer to scaled  $\omega^2$. a) Examples for fixed $D_e$, $V_\infty$, $b$ and varying $t_s$. 
b) $t_s<0$. Generation of an oscillating pattern shown by considering three different drivings with the same initial and maximum values, but progressing with different rates and over distinct time windows.
c) $t_s<0$. Turning on at the initial time faster varying parametric drivings gives rise to greater fluctuations.
}
\label{figure sigma caso 1}
\end{figure}

Curves in Figure  \ref{figure sigma caso 1} summarize the typical behavior for $\sigma$ in this case  upon superimposition of the initial 
conditions  (\ref{eq21}). For \underline{$t_s=0$},  solutions to the Ermakov equation tend to increase their value during the course of evolution  in a way similar to 
the brown curve in Fig.  \ref{figure sigma caso 1}.a). When  \underline{$t_s\neq0$} the resulting horizontal translations of the parametric frequency shape  deform 
solutions $\sigma$ in ways that may lead to significantly different curves even if the changes in the shift parameter $t_s$ appear commensurable. 
Besides the obvious remark in respect to the quantitative change in the initial value of $\sigma$, from the analytical point of view it is to point out that evolution 
now would no longer start  at an instant when the time-differential of the frequency parameter vanishes. This circumstance plays a role in the manner the system 
dynamics deviates from comportment it would display if $t_s=0$.   With the increasing of $\underline{t_s>0}$ the time interval where dynamics occur 
diminishes while the initial value $\sigma_0$ increases, and the initial slowly varying portion of $\sigma$  shrinks,  Fig. \ref{figure sigma caso 1}.a).   
Moving to the case $\underline{t_s<0}$,  the curve  $\omega(t)$ grows until it reaches its maximum value at  $t=-t_s$ to later decrease until its vanishing at  
$t_{+}$, Eq. (\ref{root tm}). If the time interval $[0,-t_s)$ is sufficiently wide,  the initial value of $\omega$   cuts  down meaningfully. 
As  Figures \ref{figure sigma caso 1}.a) and  \ref{figure sigma caso 1}.c) show, the effect is quite a gain for the initial value for function $\sigma$ and for the magnitude 
of fluctuations  occurring thereafter. Remark that in the plot \ref{figure sigma caso 1}.c) the $\omega$'s at initial time are very small, but never vanishing: resulting initial 
conditions for the corresponding $\sigma$ are relatively large, but in fact finite. The point is that when $t_s<0$ a sign change occurs for $\dot\omega$.
The system is thus required to  adapt itself first to a continuous positive frequency jump $\Delta_{\omega, 1}=\omega_{max}-\omega_0=|D_e-V_\infty|-\omega_0$, 
and later to  a continuous negative jump $\Delta_{\omega, 2}=\omega_{max}=|D_e-V_\infty |$. Such a driving dynamics calls for a bigger 
adaptation effort as long as $t_s$ takes more negative values, and  this happening is communicated through enhanced oscillations for  $\sigma$. 
Also note that when $t_s<0$ the mean value of $\sigma$ tends to increase its value during the course of evolution after a transient time where it decreases. Such dynamics is related to the occurrence of a sign change for $\dot{\omega}$. 
\begin{figure}[h]
\begin{center}
 \includegraphics[height=4.1 cm]{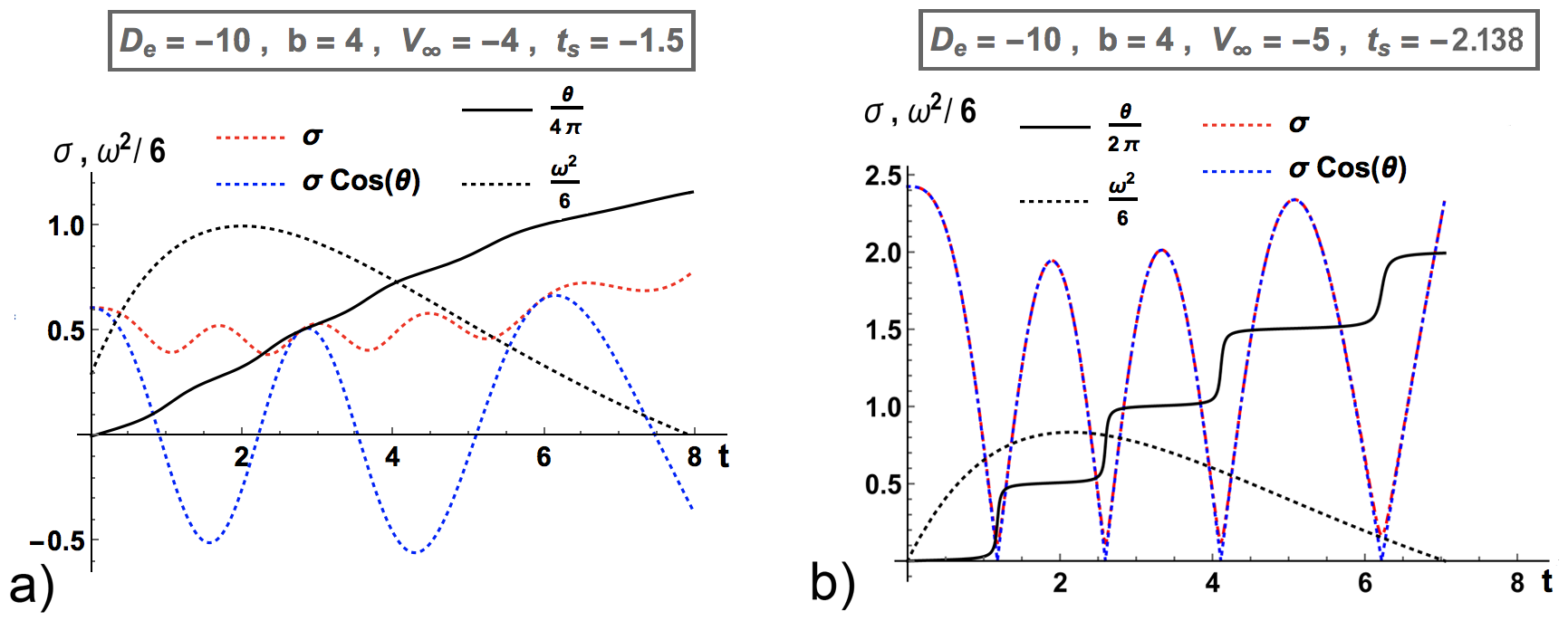} \quad \,\,
\end{center}
\caption{Comparison between solutions to the Ermakov equation $\sigma$ and solutions to the parametric oscillator equation. When its  oscillations  increase due to non-adiabaticity of frequency changes and of dynamics,     amplitude $\sigma$ progressively acquires features of solutions to the parametric oscillator equation, 
 approaching the positive portions of  $x=\sigma \cos \theta$ and the reflections across the time axes of negative ones. Accordingly, phases $\theta$ develop  continuous jumps.}
\label{figure 4 scalini}
\end{figure}
Indeed, there is a basic general process by which variations of amplitude $\sigma$ are brought about:
the raising of frequency parameter $\omega$ acts to reduce  $\sigma$ and vice versa. If the frequency parameter varies adiabatically the correspondence is 
stronger: the lowest order term produced by application of WKBJ expansion method 
entails just a 1-to-1 map  between parametric frequency $\omega$ and quasi normal-mode amplitude $\sigma$,
the relationship  $\dot{\theta}=\sqrt{c}/ \sigma^2$  between amplitude and phase of quasi-normal modes effectively turning into $\dot{\theta}\simeq \omega$. 
By plotting $\omega^2 \sigma^4 /c$ when  $t_s\geq0$ is either vanishing  or relatively small,  one would see 
 this function close to unity for most of the evolution before rapidly collapsing  to zero when  the driving $\omega$ is going to be absolutely switched off;  
 for higher values of the time-translation parameter $t_s$, the function $\sigma$ would clearly reach the fast growth regime in a shorter time. 
With the decreasing of $t_s<0$, prior the final vanishing of $\omega^2 \sigma^4 /c$ one would instead view the onset and the escalating of net oscillations, 
with meaningful  deviations from the adiabatic condition. That being the case, non-negligible fluctuations start to be generated for $\sigma$,   
with the progressive drawing near of solutions to the Ermakov equations and the absolute value of solutions $\sigma \cos\theta$ 
to the parametric oscillator equation with the same initial conditions and frequency; see Figure (\ref{figure 4 scalini}).   
A staircase type shape ensues for the phase $\theta$ with raisers developing  in correspondence of minima for $\sigma$. Owing to the development of continuous phase 
jumps  of the order of $\pi$, the function $\cos\theta$ modifies to a sequence of square-wave like elements with minimum and maximum about the values -1 and 1. 

Before to conclude the Subsection, we finally call attention on the fact that for the plots we have chosen 
frequency parameters so to exclude very strong variations for drivings.  The relatively large time-scale parameter $b$ we used allows to better visualise the small-to-large generation of fluctuations for the amplitude $\sigma$.  Lowering $b$, the driving changes  become faster, e.g.  making a more peaked shape for $\omega(t)$ about the maximum when $t_s<0$. Wider fluctuations can be then expected. Since also the time interval where the parametric driving acts becomes smaller, only a part of the previously evidenced fluctuating patterns for $\sigma$ may be realised, as shown in figure \ref{figure 4 scalini}.c).

\subsection{Case 2:   $D_e<0$  and $V_\infty>0$}
\label{ss 5 2}

The case we dwell upon in this Subsection differs from the previous one in that the parametric frequency tends to an asymptotic value $V_\infty >0$, allowing for a potentially infinite time evolution. After a possible pulse-like driving, the system  reaches exponentially fast a dynamical regime that basically act as an harmonic oscillator that is perturbed extremely weakly by a time-dependent correction to the frequency. 

\begin{figure}[h!]
\begin{center}
\includegraphics[height=3.85 cm]{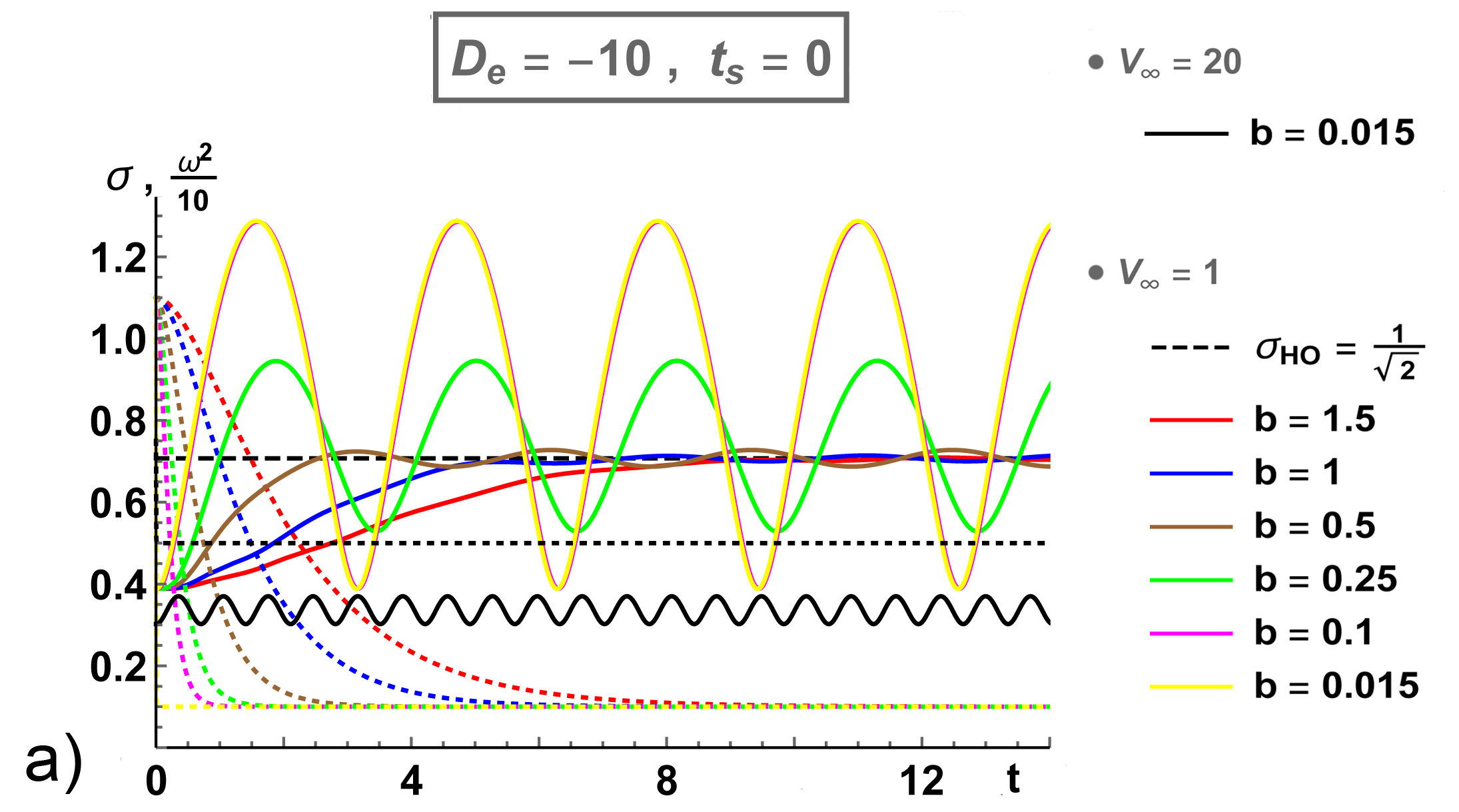} \\
\includegraphics[height=3.8 cm]{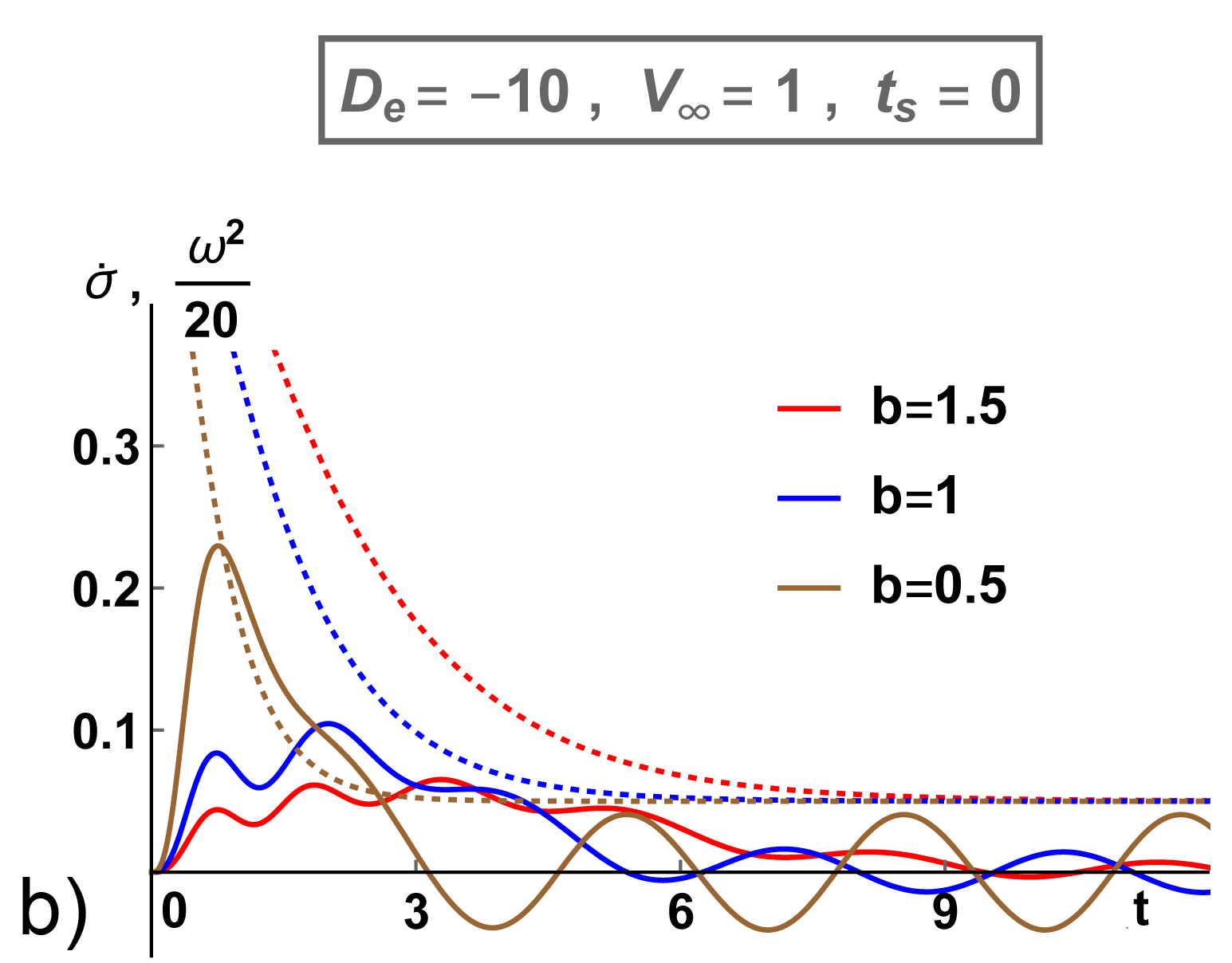} \qquad
\includegraphics[height=3.8 cm]{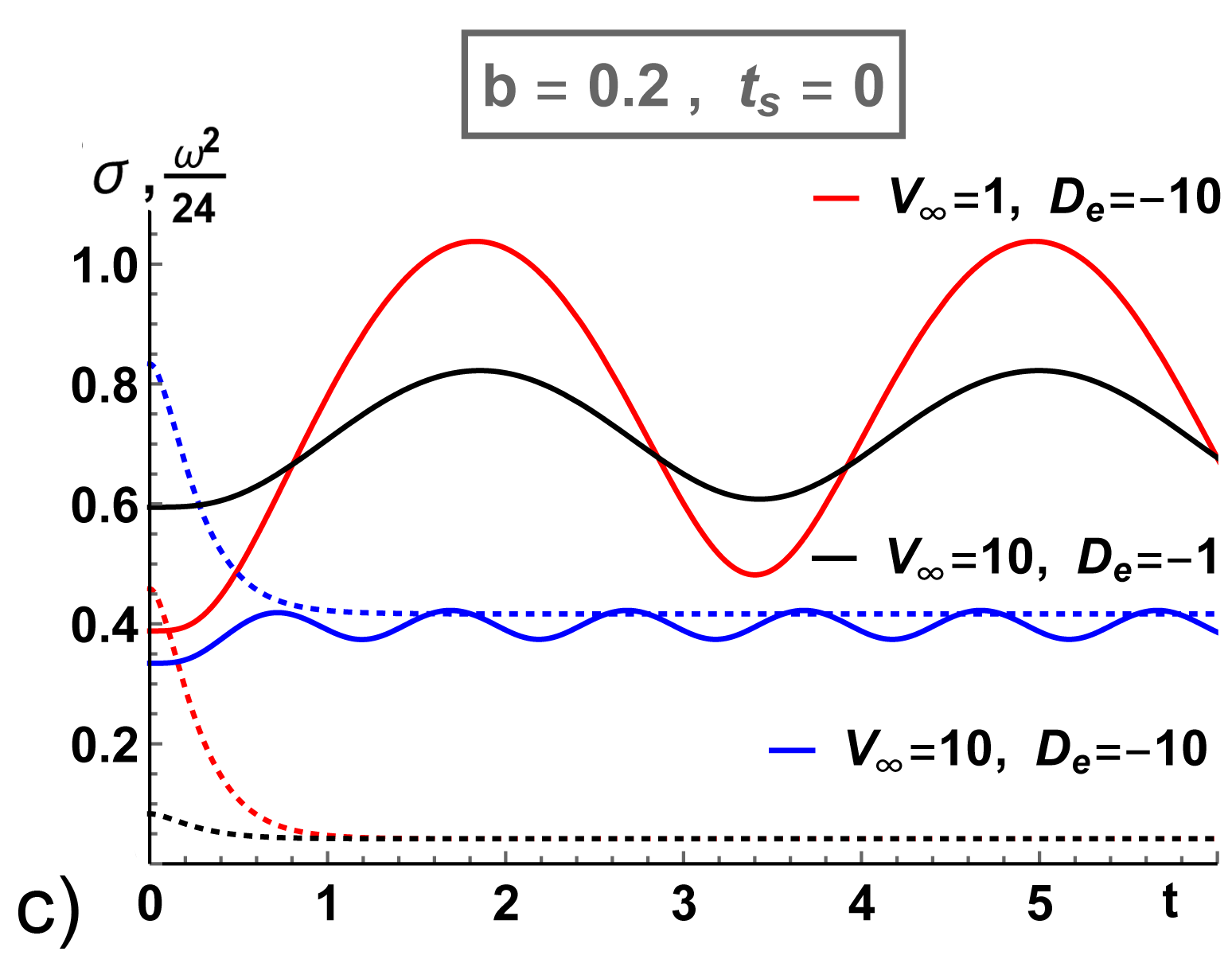}
\end{center}
\caption{Function $\sigma$ and $\dot{\sigma}$ when $V_\infty>0$, $D_e>0$, $t_s=0$ in (\ref{omega2 gen}). Initial conditions are (\ref{eq21}) with $c=1/4$. 
Color dotted curves refer to corresponding potentials up to scaling factors. Oscillations of different magnitude are generated for $\sigma$
depending on the properties of the acting parametric driving. In panel a)  the reference value $\sigma_{HO}=(c / V_\infty)^{1/4}$ is also shown for $V_\infty=1$.
 }
\label{figure sigma caso 2}
\end{figure}

Let us analyze the solutions to the Ermakov equation in the present case once  we set $\underline{t_s=0}$ and the initial conditions (\ref{eq21}). 
The frequency decreases monotonically from the initial value $\sqrt{V_\infty+|D_e|}$. 
Solution $\sigma$ initially grows until it stabilizes to an oscillating regime. For fixed  height of the frequency parameter shape (i.e. for fixed $D_e$), the magnitude of oscillation is strongly affected by the width of the  time interval  where $\dot{\omega}$ is significant and the jump for $\omega$ from maximum to asymptotic value is essentially realized. 
As long as the positive time-scaling parameter $b$ increases, the amplitude of oscillations becomes smaller and smaller, the fluctuation being basically 
negligible for sufficiently large values of $b$. 
 Magnitude of oscillations varies significantly  provided that the different $b$ are taken so to exclude 
 at a too sudden initial jump for $\omega$. This is evident from Figure \ref{figure sigma caso 2}.a) (there the $b=0.1$ and $b=0.015$ curves almost overlaps).
At the initial  stage $\sigma$ grows monotonically with very little visible fluctuations in the growth rate, as 
evidenced in Figure \ref{figure sigma caso 2}.b) where $\dot{\sigma}$ is plotted
for the smoothest  curves in Figure \ref{figure sigma caso 2}.a). At fixed $b$ and $D_e$, the greater is the asymptotic value of frequency parameter $ |V_\infty|>0$  the lower are the values about which oscillations of  $\sigma$ are realized, as well as their amplitude and their characteristic period;  see the comparison between the black curve and the (essentially overlapping) yellow and magenta ones in figure \ref{figure sigma caso 2}.a) as well as the blue curve in figure \ref{figure sigma caso 2}.c). For assigned $V_\infty$ and $b$, the progressive augment of  $|D_e|$ results  into more and more substantial departures of $\omega^2$ from the "optimal" constant value $V_\infty$, and wider oscillation regimes arise
with mean values of $\sigma$ attaining higher values with respect to the stationary amplitude  
$\sigma_{HO}=(c / V_\infty)^{1/4}$ of harmonic oscillators whose frequency is the asymptotic value $\omega_\infty= \sqrt{V_\infty}$ of $\omega(t)$; see Fig. \ref{figure sigma caso 2}.c).

\begin{figure}[h!]
\begin{center}
\includegraphics[height=3.8 cm]{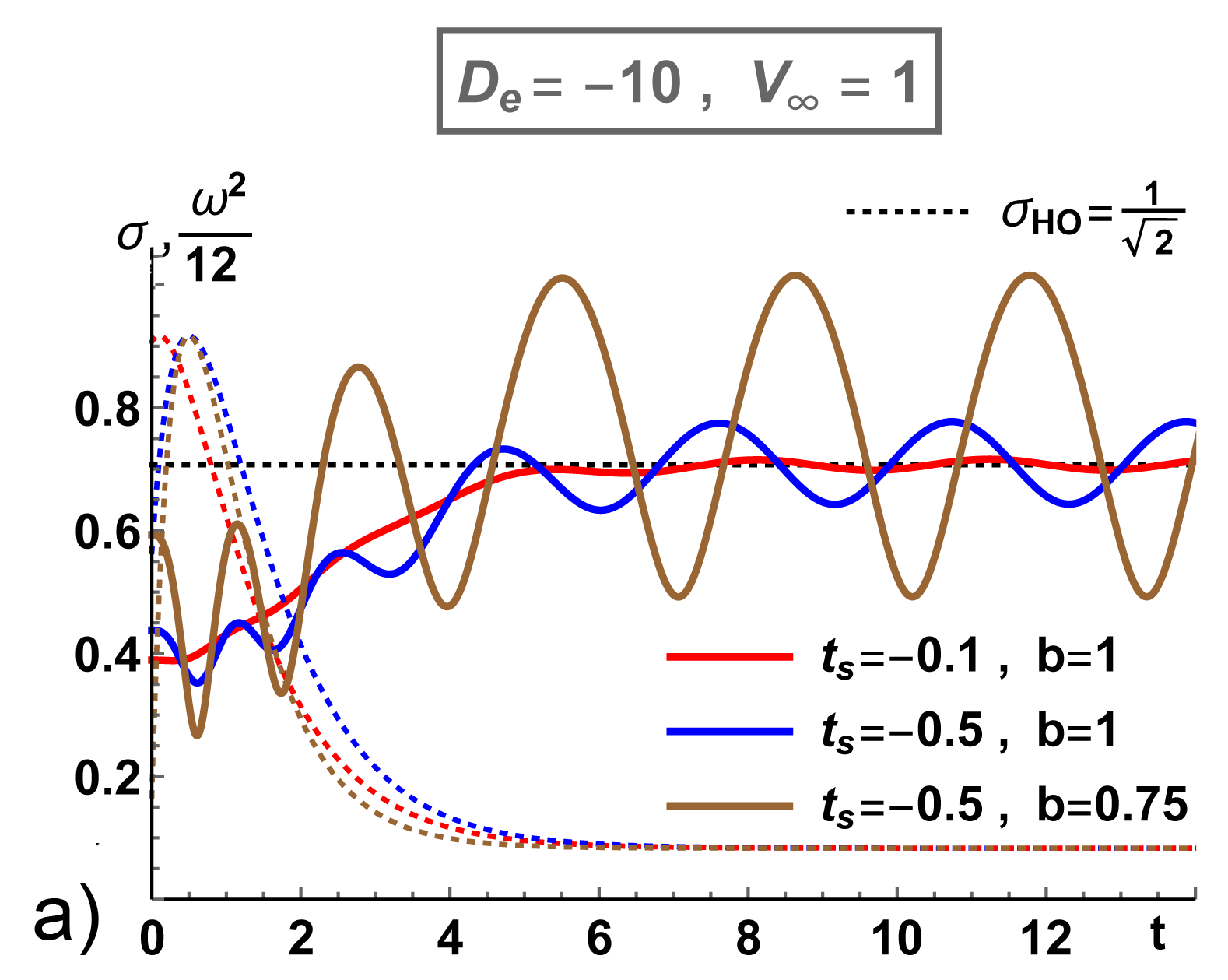} \quad
\includegraphics[height=3.8 cm]{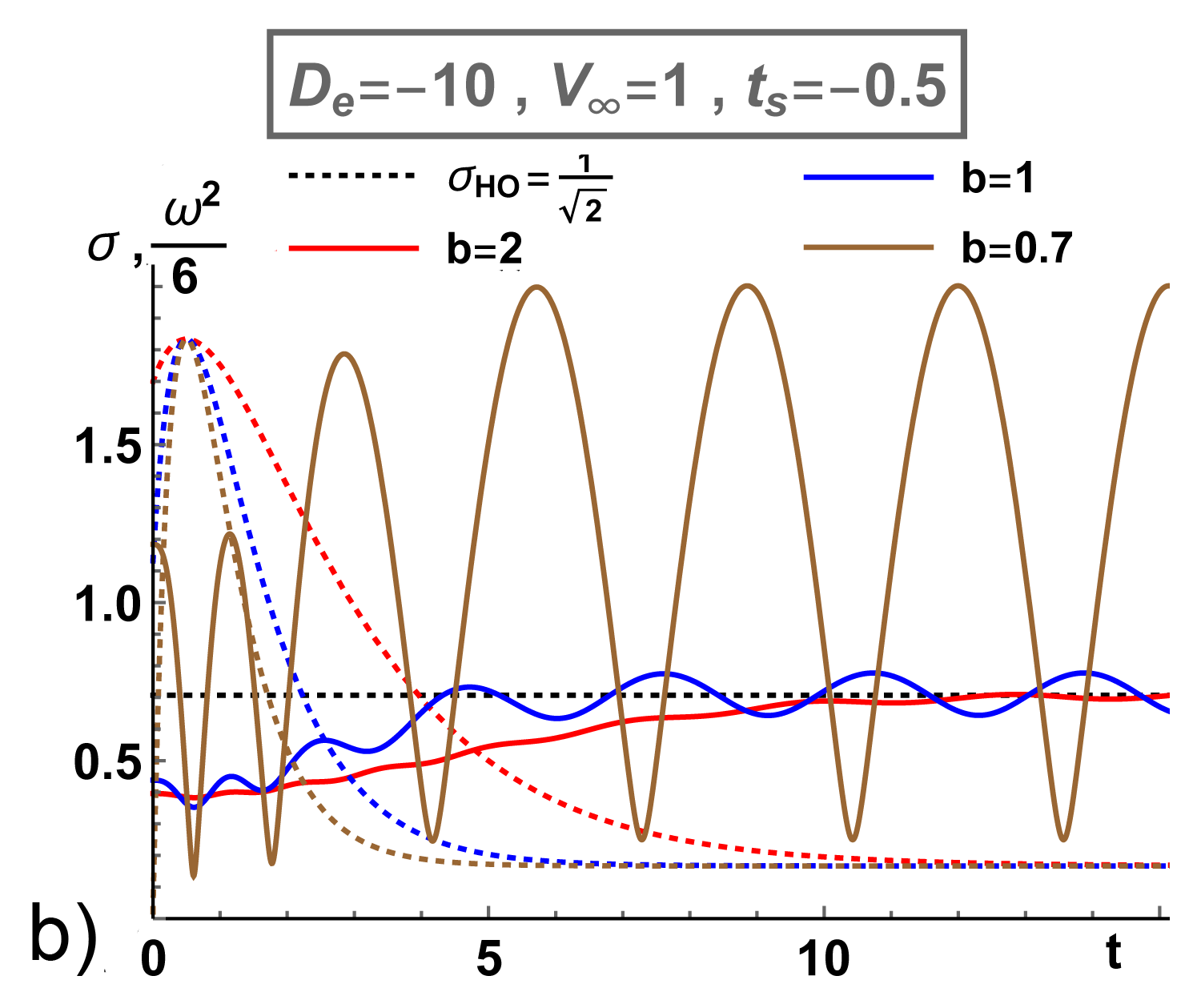} \quad
\includegraphics[height=3.8 cm]{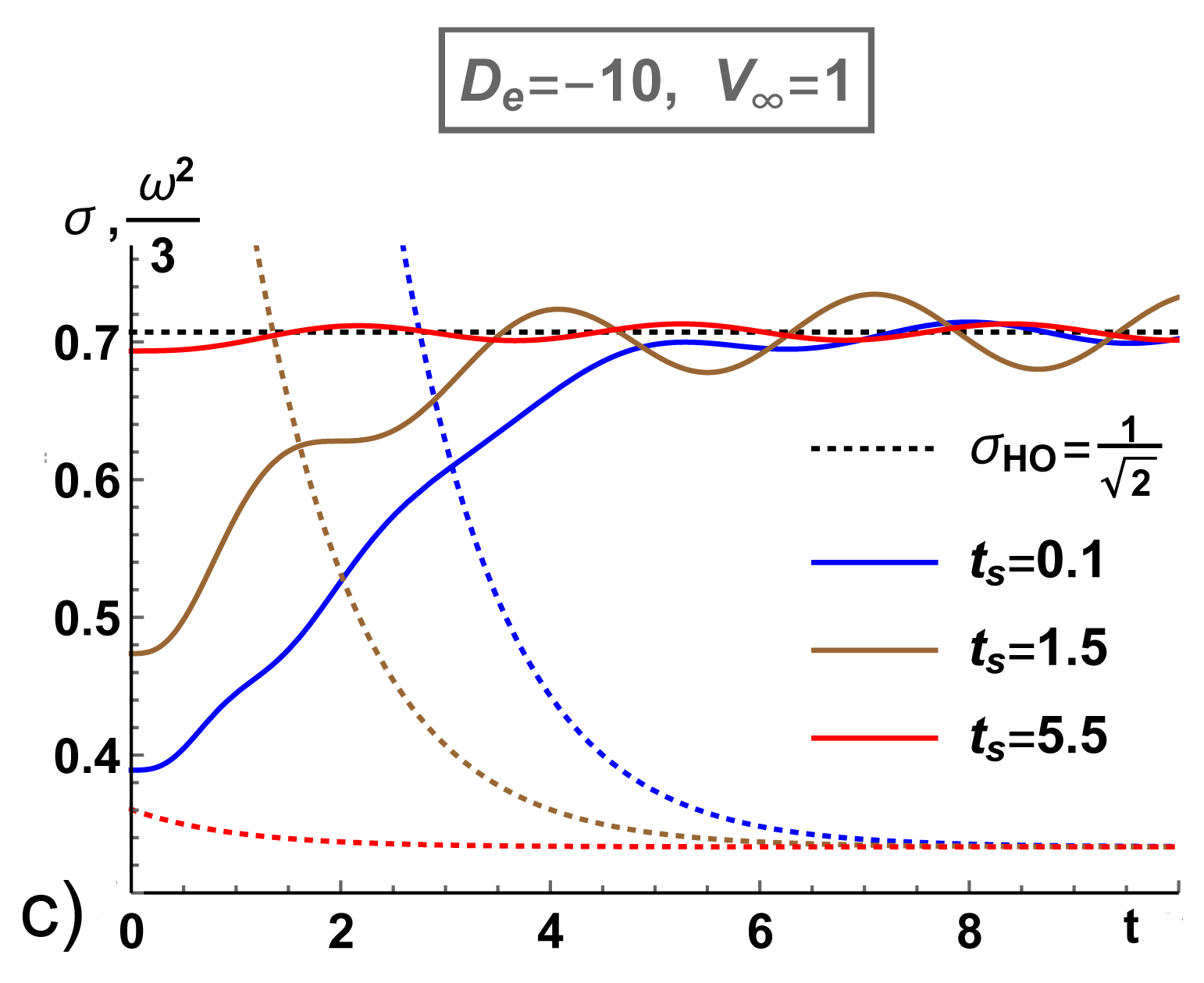} 
\end{center}
\caption{Function $ \sigma $  when time translations  are performed with $t_s\neq 0$.
Colored dotted curves refer to scaled $\omega^2$. Oscillations of different magnitude are generated depending on the properties of the acting parametric driving.}
\label{sigma_caso2_tsneq}
\end{figure}
We can now proceed by considering a nonvanishing time-translation parameter $ t_s> - b \log (1+\sqrt{1-V_{\infty}/D_e})$. 
By considering negative $t_s$ and the initial conditions (\ref{eq21}), the driving frequency turns from monotonic to unimodal in a manner that is, evidently, similar to the Case 1 argued in previous Subsection. 
By progressively diminishing the parameter $t_s<0 $, the newly introduced initial increasing  behavior for $\omega^2$ enters more and more effectively into action in a way that is in 
in agreement with findings from Subsection \ref{ss 4 1} and expectations from the WKBJ/adiabatic arguments. 
From Figures \ref{sigma_caso2_tsneq} one infers indeed that  larger fluctuations are generated for lower negative values of $t_s$. 
Explicit examples are also provided concerning the repercussion of diversifying the time-scaling parameter $b$, as in plots \ref{sigma_caso2_tsneq}.a) and 
 \ref{sigma_caso2_tsneq}.b) that make definite the resulting amplification of $\sigma$-oscillations (compare with Figs. \ref{figure sigma caso 2}.a) and \ref{figure sigma caso 2}.b)). 
Once again, it is found that that mere adiabaticity of frequency changes is not enough to ensure a smooth outcome for $\sigma$. In contrast,  
if very gentle modifications of parametric frequency begin to rule the dynamics at a certain time, then oscillations of large magnitude keeps being sustained.  This 
simply reflects the fact that the systems enters into that dynamical regime with a largely oscillating amplitude $\sigma$ owing the features of the initial transient regime, 
and the very adiabatic changes of hamiltonian taking place thereinafter are so  extremely slow that they are not much effective in altering the evolution trend that the 
amplitude $\sigma$ is already experiencing. That is, the system is led in a relatively short time and exponentially fast to a dynamic regime governed by a hamiltonian that 
is essentially  harmonic, with wide amplitudes  surviving.  If one looks at the envelope profile surrounding oscillations,  its modifications are just as adiabatic as expected
 by comparison with the parametric frequency $\omega$.
 
The case $\underline{t_s>0}$  can be finally commented by noticing that as long as $t_s$ begins to increase the oscillations tend to become more extensive, but 
 for sufficiently large $t_s$ the opposite  trend is engaged owing to the progressive  diminishing of $\omega_0^2$ (which gets closer to $V_\infty$)  and of  $\dot{\omega}$.

\subsection{Case 3: $D_e, V_\infty>0$ }
\label{ss 5 3}

We now consider the standard Morse potential with a positive vertical shift, that is (\ref{omega2 gen}) with both $D_e$ and $V_\infty$ positive  constants, and such 
to guaranteeing that also the minimum $\omega_{min}^2$ of $\omega^2$  is positive: $0<D_e<V_\infty$ when $t_s\leq0$, otherwise  $0<D_e<V_\infty (2 e^{-t_s/b}- e^{-2t_s/b}) $. 
 The case is  a dual counterpart to Case 2, to which it is linked by a reflection of the $\omega^2$ shape about the horizontal line identified in terms of the asymptotic value $V_\infty$. 
There is however a difference that has to be accounted for: at fixed $V_\infty$ and $D_e$, quite  fast initial variations of parametric frequency may happen when $t_s<0$. 
Indeed, $\omega^2$ starts from an initial value $\omega_0^2$ which, by decreasing $t_s<0$, first keeps itself comparable to the minimum value attained by $\omega^2$ for small $t_s<0$, then 
approaches the other reference value  $\omega_{\infty}^2$, and later can become arbitrarily great. 
Magnitude of oscillations for $\sigma$ is hence expected to not follow an uniform trend while right-translating the curve for $\omega$ 
 from the $t_s=0$ position. 
  
\begin{figure}[h]
\begin{center}
\includegraphics[height=3.8 cm]{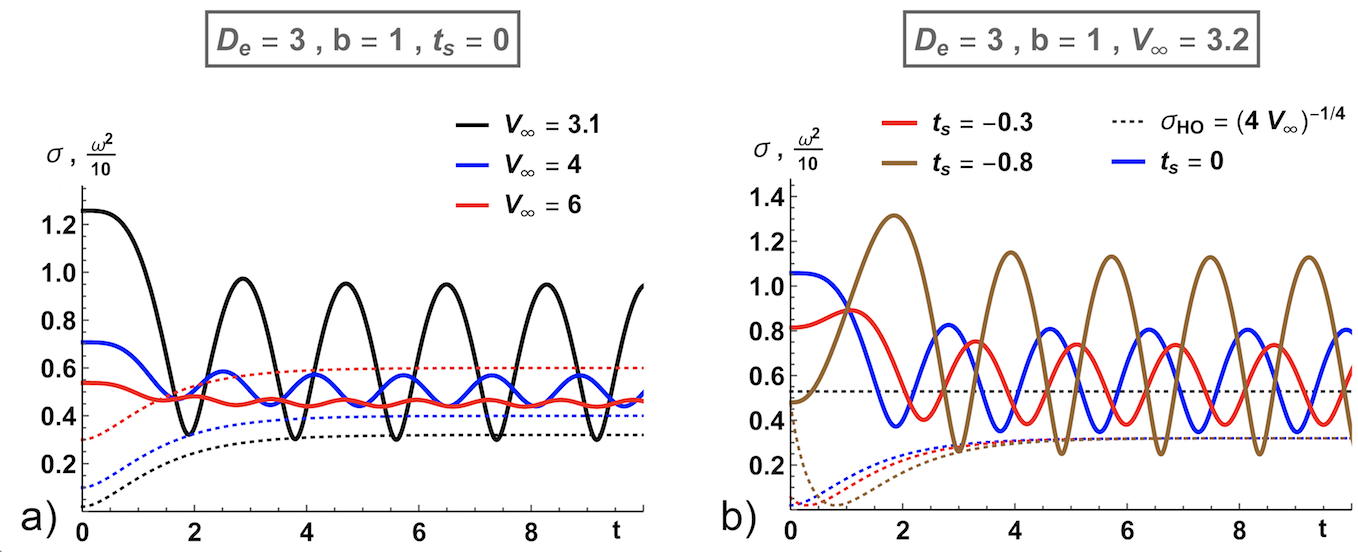} 
\includegraphics[height=3.8 cm]{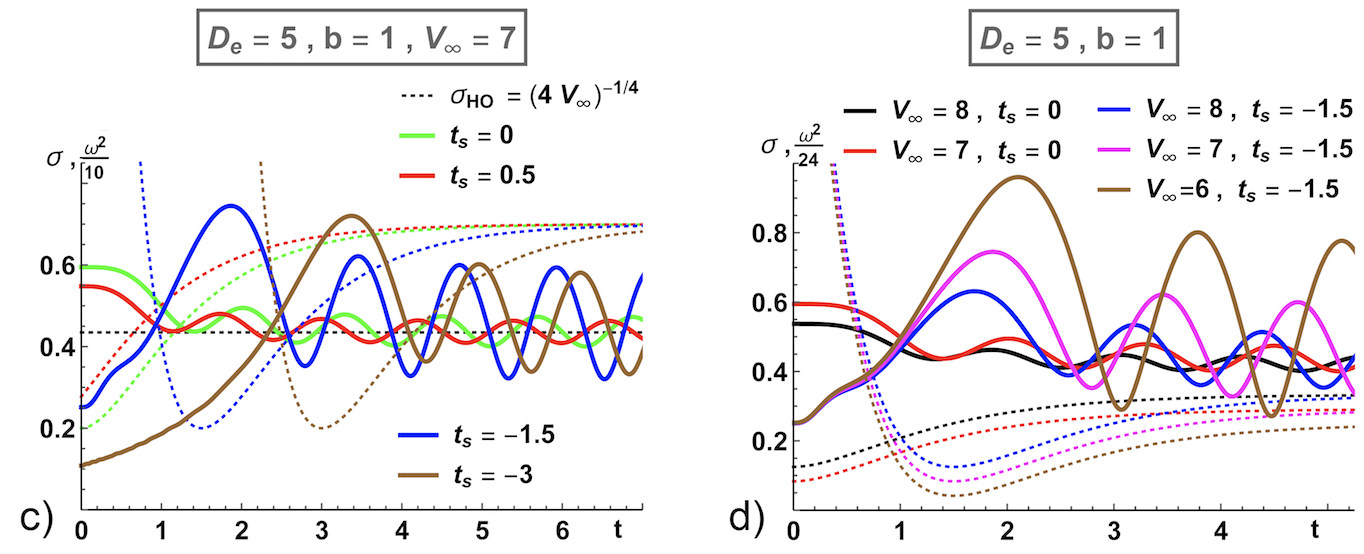} 
\end{center}
 \caption{Examples of amplitude $\sigma$ for $D_e>0$ and $V_\infty>0$. Initial conditions are (\ref{eq21}) with $c=1/4$. Colored dotted curves refer to corresponding scaled $\omega^2$.  Oscillations of different magnitude are generated by variation of the parameters  determining the time-dependent  frequency  driving.}
\label{figure sigma caso 3}
\end{figure}

Let us consider the case $\underline{t_s=0}$. The increasing of $\omega^2$ from its initial value $V_\infty-D_e$ means  that some energy is pumped into the system, and the evolution of quantities of dynamical interest reflects the presence of such basic mechanism.
When $\underline{t_s=0}$ the function $\sigma$ exhibits an asymptotic oscillating shape that is more and more suppressed as long as the difference $V_\infty-D_e$ increases, the average values also lowering; see Fig. \ref{figure sigma caso 3}.a).  Fluctuations are of greater magnitude whenever $V_\infty$ and $D_e$ are 
comparable. This happens because of the higher value of the initial value for $\sigma$ along with the wider gap between the initial minimum  and the asymptotic value for $\omega^2$, the latter giving rise to a less adiabatic dynamics. For sufficiently great values of $V_\infty$, the curve obtained for  $\sigma$ essentially flattens, as consequence of the progressive disappearance of the initial hole and of valuable rate changes for $\omega^2$.

The picture clearly modifies   if the $\omega^2$ shape previously ruling the dynamics is shifted either to the right or to the left in the $(\omega^2,t)$ plane, 
Figs.  \ref{figure sigma caso 3}.b) and   \ref{figure sigma caso 3}.c). Consider first the case  $\underline{t_s<0}$.  With the increasing of $|t_s|$ keeping  $D_e,b,V_\infty$ fixed, 
the initial value of function $\sigma$ diminishes, lessening the amplitude of oscillations maintained in the course of evolution. It is pretty clear, however, that distinguished 
situations can be expected depending on the magnitude of $t_s$. Indeed, for sufficiently small $t_s<0$, the overall qualitative dynamics tailors pretty much to the $t_s=0$ case:  
the initial decreasing trait introduced for the parametric frequency  only acts on quite a short time window and the difference between curves' fluctuations at later times 
(i.e. when $\omega^2$ approaches its asymptotic value $V_\infty$) is slightly reduced by comparison with the commencing gap $\sigma(-t_s)-\sigma_0$ (recall that now 
the minimum $\omega_{min}^2=V_\infty-D_e$ for $\omega^2$ is at $t=-t_s>0$). While lowering further the value of $t_s$, the first part of the amplitude curve goes down and larger fluctuations appear. So, for sufficiently small negative $t_s$ the  
magnitude  of $\sigma$ fluctuations  reduces compared to the $t_s=0$ case, and it  increases as expected only later  for smaller negative $t_s$. 
Besides, we checked that,  by diminishing  $t_s$ further, fluctuations invert few times again the tendency to vary magnitude monotonically, until for sufficiently low $t_s$  the 
 part of $\sigma$ with regular oscillations  appear to freeze and simply translates with the almost constant part of the parametric frequency. 
 These evident changes in the behavior trends seem to be very peculiar of the case under consideration in this Subsection, for which appears a different progression of the competing effects between the jumps taking place between the initial, the estremal and the asymptotic values attained by frequency by monotonic changes of the driving parameters. 
The  waxing and waning  mechanism way can indeed be easily checked out by looking how orbits in space $(\sigma,\dot{\sigma})$ would flow due to changes of the initial conditions  implicated for the Ermakov equation.

Vertical translation of the $\omega^2$ curve towards the time axis augments the magnitude of $\sigma$ fluctuations. Notwithstanding it may result in a minor  variation for $\omega_0^2$ (and hence $\sigma_0$), a change in the asymptotic value $V_\infty$ importantly  alters  the evolution of solutions to the Ermakov equation for quasi normal amplitude  $\sigma$.  
Figures  \ref{figure sigma caso 3} show that the effect is really worthy of attention. 
 For instance, in examples shown in graphics \ref{figure sigma caso 3}.d) with $t_s<0$, the asymptotic value $V_\infty=7$ is translated by  a factor that is only  $1/7$ of its value, but
the relative changes in magnitude of oscillations for $\sigma$ are greater as a consequence of having either halved or increased by $50 \%$ the  jump $\omega^2_\infty-\omega^2_{min}$ 
(initial decreasing traits for the parametric frequency and jumps $\omega^2_0-\omega^2_{min}$ are on equal footing because the 
$\omega_0$ experience a percentage variation that are less than $1 \%$).
Figures \ref{figure sigma caso 3}.b)-\ref{figure sigma caso 3}.d) clearly indicate also that oscillations 
develop, once again, not symmetrically with respect to the reference amplitude $\sigma_{HO}=(c / V_\infty)^{1/4}$ associated with the harmonic oscillator whose frequency is 
the asymptotic value  $\omega_\infty=\sqrt{V_\infty}$ of parametric frequency $\omega(t)$.  What happens instead when the time shift parameter  attains positive values 
$\underline{t_s>0}$ is very comprehensive: the hole for $\omega^2$  tends to progressively disappear and the initial value $\omega^2_0$ assumes higher values,
 thereby implying for $\sigma$ that oscillations of smaller magnitude are realized about $\sigma_{HO}$.

Figures \ref{figure sigma caso 3}, as well as Figs.  \ref{figure sigma caso 2} provided dealing with Case 2,  show an evident consistency with the Sturm theorem, 
in that the curves $\sigma$ resulting for majorant $\omega^2$ possess  (in a sufficiently large time interval) more extrema.  
Even though the concrete examples we  explicitly reported by specifying numerical values for the parameters involved are such that the resulting parametric frequency 
$\omega^2$ are relatively similar and the effect may not be marked, they still reveal that the frequency of occurrence for the maxima/minima exhibited by the quasi-normal 
mode amplitude $\sigma$ does respond to the introduction of majoring/minoring the pumping mechanism for the parametric oscillator (\ref{eq27}). 
In figure \ref{sigma_caso2_tsneq}.a), for instance, effects of the different number of maxima for $A x_1^2+C x_2^2+2Bx_1x_2$ in Eq. (\ref{eq30}) 
are already visible in few time units through an increasing dephasing that is realized. The dynamics of the harmonic oscillator with the minimum value attained by 
parametric frequency introduces some limits on the oscillation mechanism for the solution to the parametric oscillator equation with the Morse-based frequency 
(\ref{omega2 gen}). From the Sturm comparison theorem  follows indeed that a zero for the parametric equation with the frequency (\ref{omega2 gen}) occurs 
between two successive zeros of solutions to the harmonic oscillator with the asymptotic frequency  $\omega_\infty=\sqrt{V_\infty}$  (Case 2) or 
$\omega_{min}=\sqrt{V_\infty-D_e}$ (Case 3). 

We conclude this Subsection by calling attention of the fact that  variations of $b$ has not explicitly argued since  the issue can be in effect concluded 
on the grounds of what we have  understood previously.

 \subsection{Case 4: $D_e>0$ and $V_\infty<0$}
 \label{ss 5 4}
 
 The Morse-type structure (\ref{omega2 gen}) can be finally considered with $D_e>0$, and
  with a negative vertical shift $V_\infty<0$. 
 In order to avoid $\omega^2<0$, the case has to be concerned with suitable negative shifts of time variable $  t_s < -b \log(1+\sqrt{1-V_\infty /D_e})$ and dynamics over finite time intervals.  The case regards  therefore a parametric frequency  starting from a positive value at initial time, and then monotonically decreasing very rapidly
until it vanishes at a certain time (recall Figure \ref{figura morse} and  Footnote 1 at page 7).
 Inspection of  the independent solution to the parametric oscillator equation shows a possible generation of initial zeroes and  a rapid transition to the exponential growth regime.
Once one looks at function $\sigma(t)$, the evolutionary trend would appear rather similar to what we have seen in Case 1 with $t_s>0$, with quite modest undulations and the
maximum reached at the end of evolution. The dissimilar degree of  non-adiabaticity  in the monotonic decreasing of the driving $\omega$ for the two cases intervenes 
at quantitative level. The changes of the parameter $b$ determining the time-scale for evolution have an impact  on the frequency and its decreasing rate for the present case; 
 indeed, while the duration of signal is affected relatively little,  differences for the initial value $\omega_0$   can be noteworthy.
The initial value for the amplitude function defined via Eq. (\ref{eq21}) substantially cuts down accordingly.

\section{Quantum variances and uncertainties}
\label{ss 6}

\subsection{Classical and quantum squeezing of orbits}

Solutions to equation of motion for the parametric oscillator define curves in the extended phase space
 $(q_c,p_c,t)$, where $q_c$ and $p_c$ denote the classical position and momentum dynamical variables.   
 The collection of all possible orbits in such space that are compatible with an assigned value $I_{cl}$ of the classical Ermakov invariant
(i.e., Eq. (\ref{eq8}) with $q\to q_c$ and $p\to p_c$) spans a surface that has the topology of a cylinder.
This surface is essentially determined by the classical Ermakov invariant $I_{cl}$  which, 
at any given time, defines an elliptic curve in the position-momentum space that collects all the possible pairs 
$(q_c,p_c)$ obtained by variation of initial conditions for the Ermakov amplitude equation. 
 During the course of evolution the invariant $I_{cl}$ generates different ellipses about the origin in the plane $(q_c,p_c)$  because the functions 
 $\sigma$ and $\dot{\sigma}$ vary. 
These ellipses are of fixed area $\frac{\sqrt{c}\, \pi}{ 16 m}=\frac{\pi m I_{cl}}{\sqrt{c}}$ but with foci placement, eccentricity, axes, and length changing in time. 
Their motion can be schematically be ascribed  to variations of quantities $\sigma$ and  $\dot{\sigma}$. In particular, defining 
 \beq
\delta=\frac{\sigma\dot{\sigma}}{\sqrt{c}} \,\,  
\label{def delta}
\eeq
for  convenience, intersections of ellipses with axis in phase space $(q,p)$ are identified  via 
$q_{c,\pm}^*=\pm \sqrt{I_{cl}} \,  \sigma / \sqrt{  c \,(1+\delta^2) }$ and $ p_{c,\pm}^*=\pm m \sqrt{I_{c}}/ \sigma $.
The identification of the classically accessible region in phase space put a basis for the connection between classical and quantum aspects of the parametric oscillator dynamics,  which is rather naturally validated within a coherent-states approach, where dynamical reshaping of the elliptic sets collecting all couples of position-momentum coordinates that can be realized classically at each time is conveyed at the quantum level to the Wigner ellipses
(standard Wigner ellipses  result  upon replacements $q_c\to\ q_c-\mean{q}$ and $p_c\to p_c-\mean{p}$  in the Ermakov invariant, which is set to the value $ \hbar/2m$). 
Indeed, the coefficients of the quadratic invariant $I_{cl}$ can be rephrased as second statistical momenta for the position and momentum operators. 
Precisely, by expressing these operators as linear combination of the operators $a(t)$ and $a^\dagger(t)$ through inversion of Eq. (\ref{eq10}), 
it is promptly seen that their variances over Lewis-Riesenfeld  coherent and number states, $\ket{\alpha}$ and $\ket{n}$,  take the form 
 \beqa
\Delta_\alpha q =\sqrt{ \frac{\hbar   }{2m  }}\,   \, \frac{\sigma}{c^{1/4}} \,\, ,  
\qquad 
\Delta_\alpha p =
\sqrt{\frac{\hbar \, m    }{2 }} \,\, \frac{c^{1/4}}{\sigma} \, \sqrt{1+\delta^2 } \,\ ,  
\qquad 
\Delta_n q=\sqrt{1+2n} \, \Delta_\alpha q \,\, , 
\qquad \Delta_n p=\sqrt{1+2n} \, \Delta_\alpha p \,\, ,
\label{varianza p alpha} 
\eeqa
 yielding to the Heisenberg uncertainties $\Delta_{\alpha}q\,\, \Delta_{\alpha} p=\frac{\hbar}{2}  \sqrt{1+\delta^2}$ and 
  $\Delta_n q\, \Delta_n p=\left(n+\frac{1}{2}\right) \, \hbar \, \sqrt{1+\delta^2}$.
The quantity $\hbar \delta=\hbar \sigma\dot{\sigma}/\sqrt{c}$ can be thus recognized as twice the position-momentum correlation among Lewis-Riesenfeld coherent states 
$\ket{\alpha}$, $\bra{\alpha} qp+pq\ket{\alpha} -2 \bra{\alpha} q\ket{\alpha} \bra{\alpha} p\ket{\alpha}=\hbar\,\delta$, and as $2 (1+2n)^{-1}$ times the position-momentum 
correlation among Lewis-Riesenfeld number states. 

 Having already scrutinized the behavior of the amplitude $\sigma$, to complete gaining insight on the transfer at quantum level of the  squeezing dynamics for classical Ermakov ellipses we shall provide examples  that outcome for the function 
\beq
\Delta_{UP}=\frac{\hbar}{2}\,\, \left( \sqrt{1+  \delta^2} -1 \right) \,\, 
\label{eq Delta UP}
\eeq
measuring the deviation from the stationary frequency case of the position-momentum Heisenberg uncertainty over coherent  and number states.
Results of previous Section for quasi-normal mode amplitudes $\sigma$ provide a first glimpse in  respect to possible evolutions for $\delta$ and $\Delta_{UP}$.  
Withal, we can notice that the Ermakov equation can be managed to generally write 
  $\sqrt{c} \dot{\delta}= \left( c- \omega^{2} \sigma^4 \right)    \sigma^{-2}+  \dot{\sigma}^{2} =[\left( c- \omega^{2} \sigma^4 \right) +c \, \delta^{2}] / \sigma^{2} $, 
that gives an indication  about the way the degree of adiabaticity of amplitude $\sigma$ changes can act in squeezing Ermakov-Wigner ellipses and build up nontrivial 
correlations in the course of system's evolution. We thus expect that in some cases deviation of products $\Delta_{\alpha}q\,\, \Delta_{\alpha} p$
cannot be neglected compared to the familiar minimum value $\hbar/2$.

\subsection{Position-momentum uncertainties. Case 1: $D_e,V_\infty<0$}

\begin{figure}[h!]
\begin{center}
 \includegraphics[height=3.59 cm]{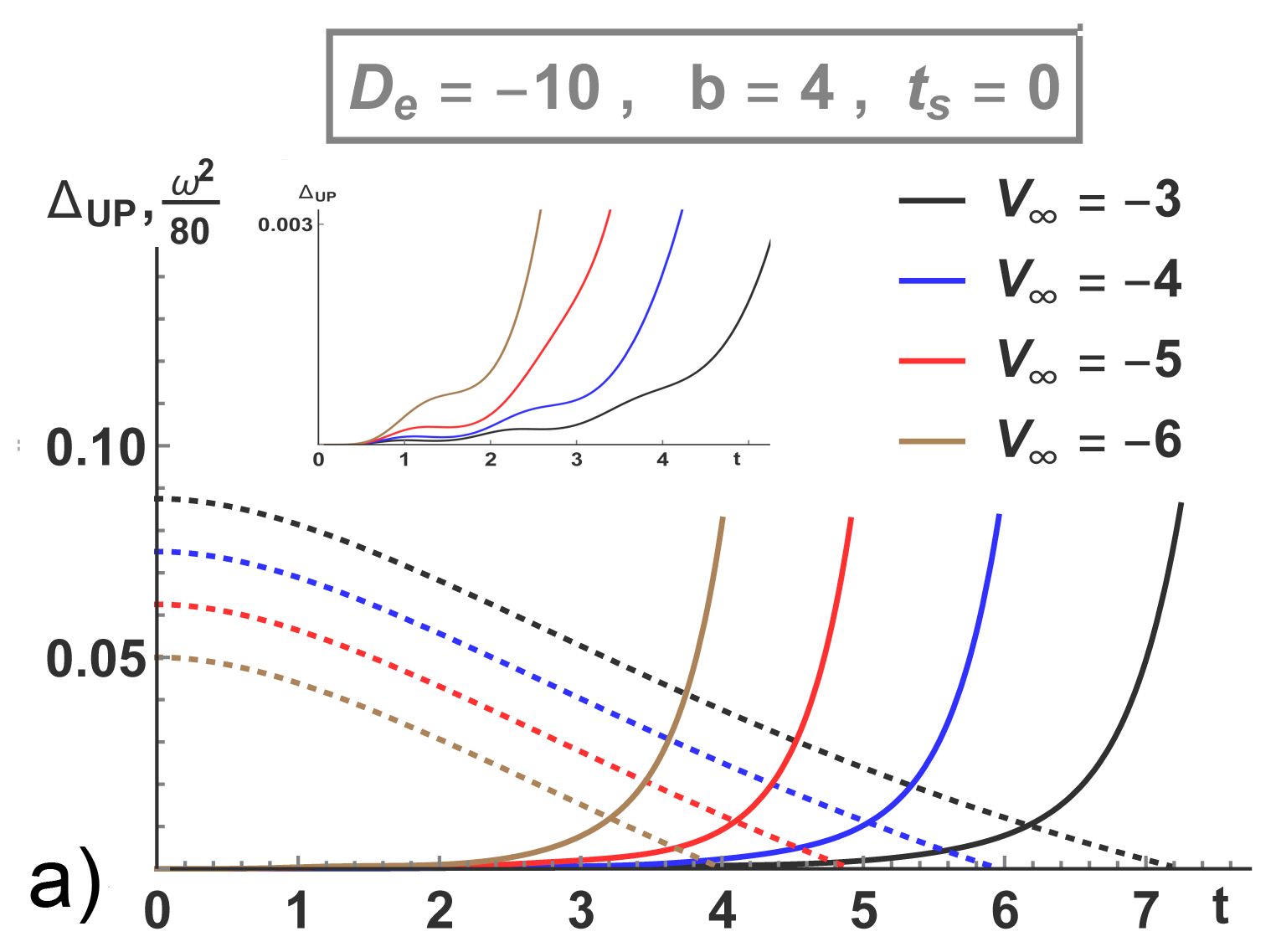} \quad
\includegraphics[height=3.6 cm]{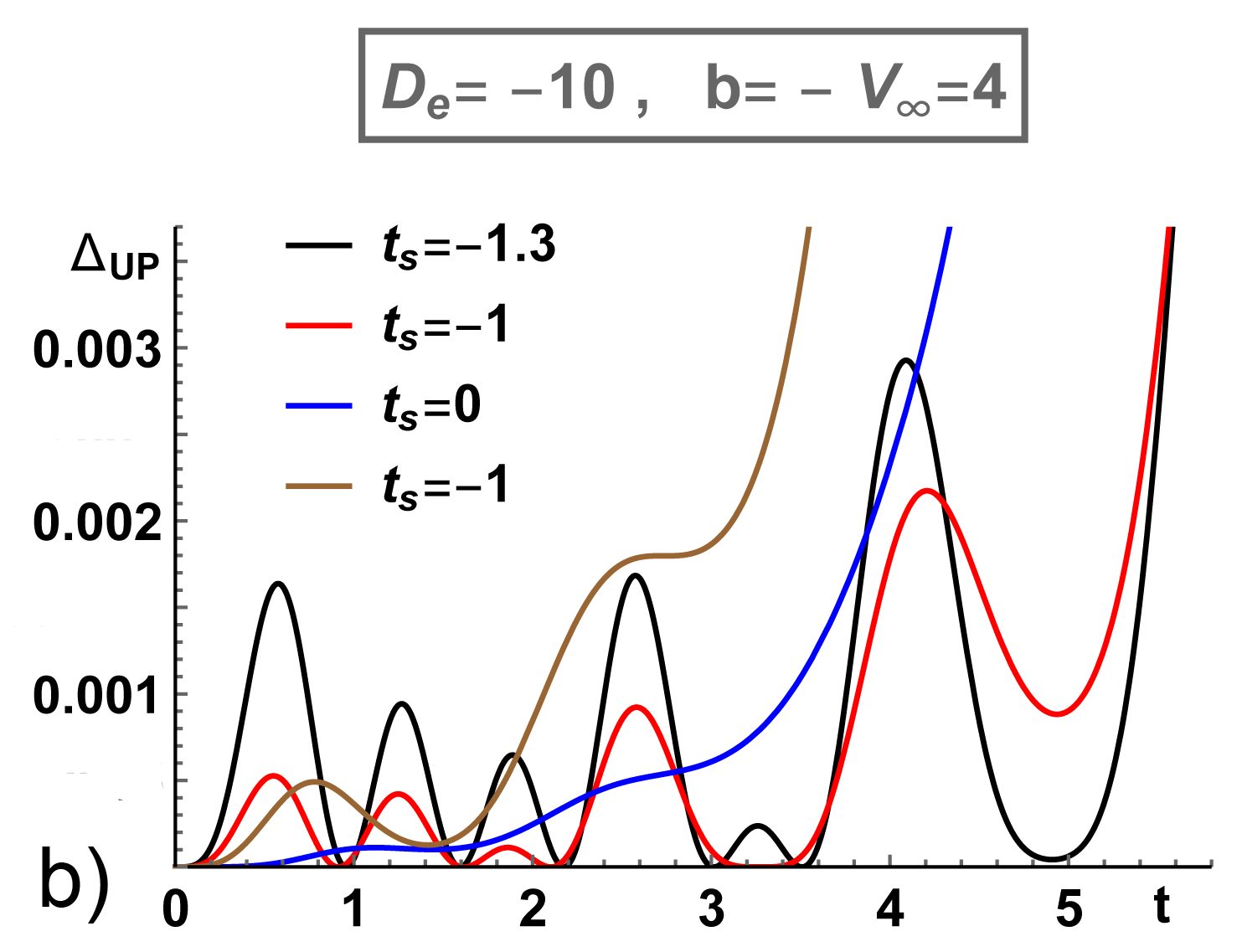}\quad
\includegraphics[height=3.6 cm]{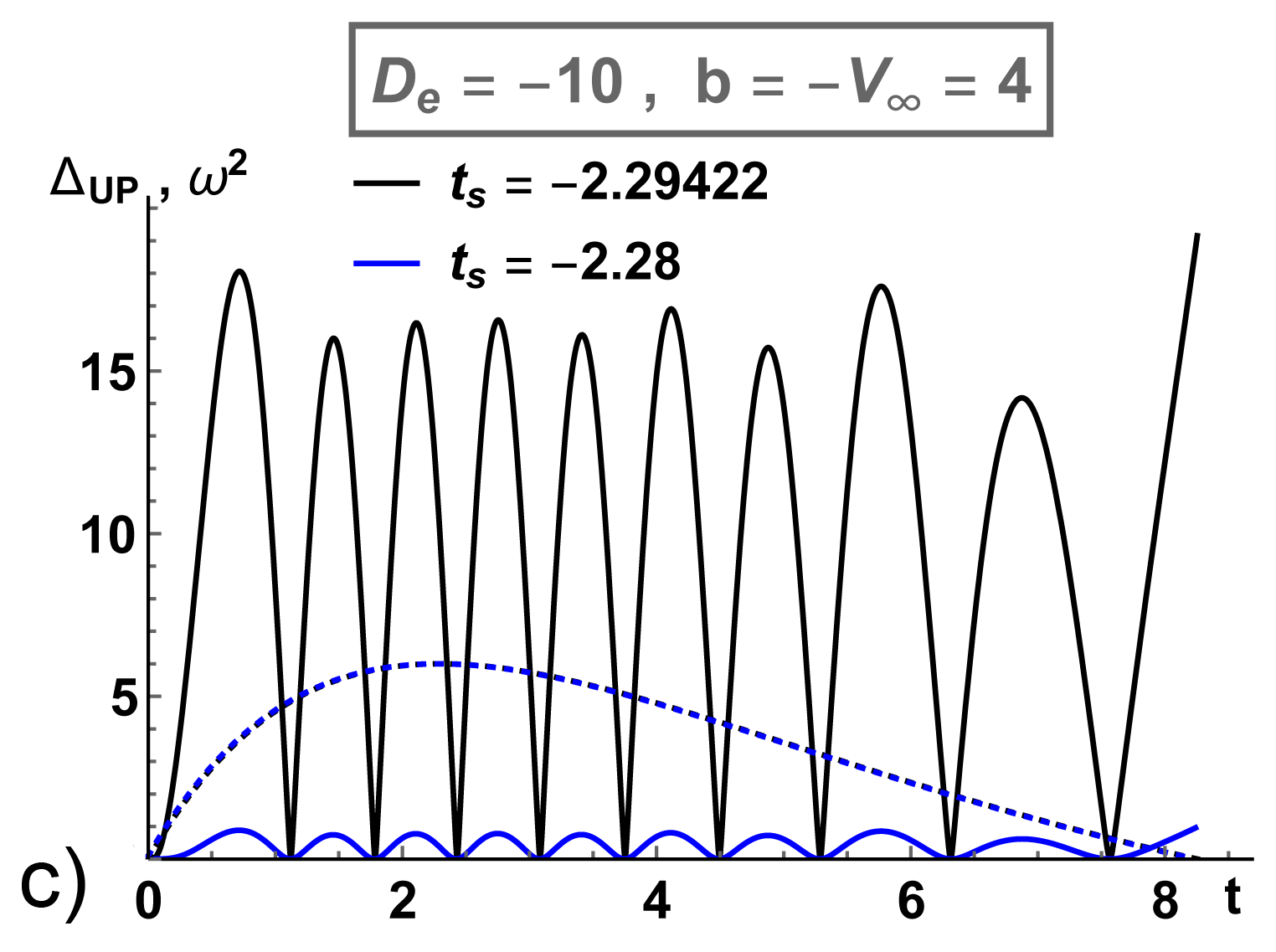}
\end{center}
\caption{ $\Delta_{UP}$  in $\hbar=1$ units
 when 
 $\omega(t)$ is given via  Eq. (\ref{omega2 gen}) with $D_e,V_\infty<0$. a) Examples  with fixed $D_e, b$ and $t_0=0$ and  varying $V_\infty$. The plot in the upper corner gives a close-up of initial trait of the curves.  b) Initial traits of curves $\Delta_{UP}$ obtained with positive, vanishing and negative values of $t_s$. c) Generation of large fluctuations when $t_s<0$. Dotted curves   refer to  parametric frequencies generating the 
$\Delta_{UP}$ plotted with solid lines  of the same color (overlapping in last plot).   }
\label{figura UP caso 1}
\end{figure}

\indent Assuming the initial conditions (\ref{eq21}),  the quantity (\ref{eq Delta UP}) is  vanishing at initial time. Main aspects concerning  what happens after that  time are shown in Figure \ref{figura UP caso 1}. They  can be subsumed as follows. 
When $\underline{t_s=0}$  the growth of $\Delta_{UP}$ is fairly smooth and slow, before to raising markedly  in the final course of dynamics as the time (\ref{root tm}) is going to be approached. In Figure  \ref{figura UP caso 1}, for instance, the maximum gain for $\Delta_{0}\hat{q}\, \Delta_{0}\hat{p}$ is a bit below $20 \%$ of its initial value.   
 Lowering the parameter $V_\infty$, 
the pumping mechanism operates on a shorter time interval and $\Delta_{UP}$ varies faster. 
If $t_s$ is varied from zero, forewarning of oscillations begin to arise.  
 The generation of a 
 succession of peaks is distinct when $\underline{t_s<0}$, which is the case  where 
a sign change is experienced by $\dot{\omega}$.   The lowering of $t_s$ makes the oscillating pattern more uniform, but also of greater magnitude as the shape of the frequency parameter $\omega$   varies more before the maximum is reached  at  $t=-t_s$. When $\omega_0$ gets very  close to zero, minor variations of $t_s$ leading to  inappreciable variation of  the frequency parameter shape obviously provoke impressive alteration for the magnitude of oscillations; in Figure \ref{figura UP caso 1}.c), this is shown with values for the maxima of  $\Delta_{UP}$ that 
jump at great rate from about $\hbar$ to more that 15 times that value.

\subsection{Position-momentum uncertainties. Case 2:  $D_e<0$  and $V_\infty>0$}

\begin{figure}[h!]
\centering
 \includegraphics[height=3.8 cm]{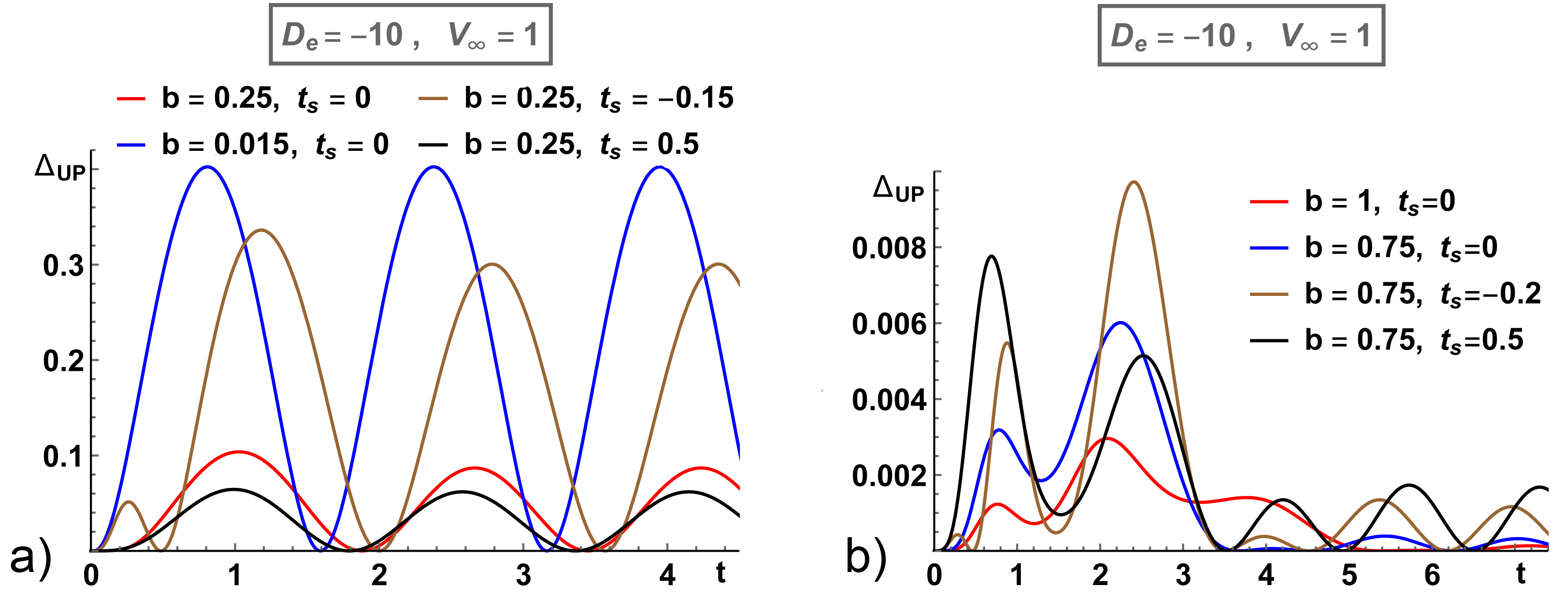} 
\caption{
Examples of $\Delta_{UP}$ curves  in $\hbar=1$ units when $V_\infty >0$  and $D_e <0$  in (\ref{omega2 gen}). $D_e$ and $V_\infty$ are hold fixed while there are two different varying sets for $b$ and $t_s$. Smaller values of $b$ amplify the deviation from minimal Heisenberg uncertainty  $\hbar/2$ and hide initial transient effects.}
\label{figura UP caso 2}
\end{figure}

Once again, the position-momentum Heisenberg uncertainty may depart pretty much  from the stationary result.
The effect may be sustained in the course of dynamics, or  first and foremost  confined at early stages when frequency changes are  more gradual.  
Figures  \ref{figura UP caso 2} are obtained by varying parameters $b$ and $t_s$.  Figure \ref{figura UP caso 2}.a) shows examples where $b$ is small and peaks 
of $\Delta_{UP}$ range from about $10\%$ to $80\%$ of the initial uncertainty value $\hbar/2$.  In Figure \ref{figura UP caso 2}.b) the parameter $b$ that scales the time 
variable $t$ is increased and magnitude of $\Delta_{UP}$ falls off consistently.  A  transient initial dynamics is clearly seen, especially for $t_s\neq0$,  where  the largest 
contributions to the position-momentum uncertainty are recorded (not exceeding $2\%$ of $\hbar/2$ with parameter used in Fig.  \ref{figura UP caso 2}.b) ).

\subsection{Position-momentum uncertainties. Case 3: $D_e>0$  and $V_\infty>0$}

\begin{figure}[h!]
\centering
\includegraphics[height=3.8 cm]{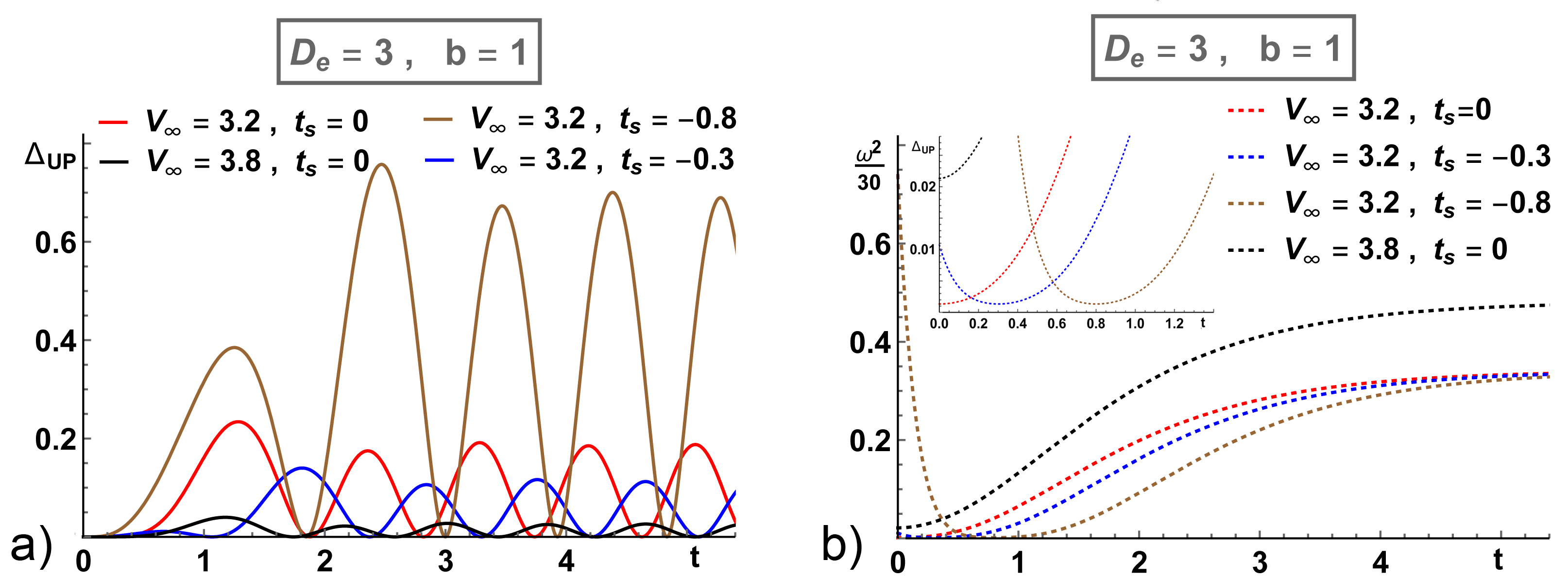} 
\caption{Case 3: $D_e, \,V_\infty>0$.
a) Oscillations generated for $\Delta_{UP}$ in $\hbar=1$ units. b) Scaled $\omega^2$ relative to examples in a). }
\label{figura UP caso 3}
\end{figure}

 Even  when both $D_e$ and $V_\infty$ in  (\ref{omega2 gen}) are positive constants, amplitude fluctuations may generate oscillating enhancements of the Heisenberg uncertainty for certain values of parameters determining the shape of frequency parameter.  Examples are given in Figure \ref{figura UP caso 3}.a) with parameters that make the effect  standing out.
The increasing of $V_\infty$ at fixed $D_e$ suppresses responses for $\Delta_{UP}$. When $t_s$  decreases from the null value, a damping of the $\Delta_{UP}$-oscillations first results that is next replaced by the foreseen amplification trend. Relatively limited reshaping of the parametric driving can result in 
evident changes for magnitude of oscillations; see red and blue curves in both panels of Figure \ref{figura UP caso 3}. As we learned Subsection \ref{ss 5 3},
the proximity to the null value of initial frequency value is indeed a crucial matter.

We do not present figures  for Case 4, for which we found curves for $\Delta_{UP}$ that resemble those of  Fig. \ref{figura UP caso 1}.a) of 
Case 1 but with a  marginal indication of an underlying fluctuation mechanism before the final fast growth.

\section{Zero-delay second order correlation functions}
\label{ss 7}

 For investigating the evolution of a quantum field governed by a time dependent quadratic hamiltonian, it is convenient to construct  either the eigenstates of the time-dependent annihilation operator or those of the time-dependent number operator,  and to connect them to the standard coherent and number states for the harmonic oscillator  identified by factorization of the hamiltonian at an initial time. 
This allows to realize a description of the dynamics based on Lewis-Riesenfeld coherent- or  number-type states, and to interpret the results in terms of stationary coherent states or 
modal excitations constructed through  a  diagonalization of the Hamiltonian at some initial time.
Features of the eigenstates of $I$ are intimately connected to those of eigenstates of the harmonic oscillator  with mass $m$ and frequency $\omega_0$ because of the extended canonical transformation linking each to the other (Section \ref{s2}),  but they present specific peculiarities induced by the concrete form of the transformation, i.e. the 
vacuum state actually considered. The probability of finding an element from the 
modal Fock basis $\{\ket{n}_0\}$ in a Lewis-Riesenfeld squeezed coherent  state $\ket{\alpha}$ does not follow a Poisson distribution.
Related to the mechanism of time-dependent squeezing of quadrature components,  
modifications of modes counting statistics are manifested during the evolution of the one-mode quantum system governed by a parametric oscillator  hamiltonian operator.  
They can be highlighted by determining the values  of the the second-order degree of coherence \cite{walls}.
In this Section we shall therefore analyze the zero-delay correlation functions  
  \beq
g_{2}=
\frac{ _{0} \bra{n} \hat{a}^{\dag \,2} \hat{a}^2 \ket{n}_0 }{_0 \bra{n} \hat{a}^\dag 
\hat{a} \ket{n}_0 ^{\,\,2}} \,\, = 
 \frac{_0  \langle n| N^2 |n \rangle_0\,\,-\,\,_0 \langle n| N |n\rangle_0}{_0 \langle n|N |n\rangle_0^{\,\,2}} \,\, .
\label{def g2}
\eeq 
The function $g_2$ can be determined on purely algebraic grounds on account of the Bogolubov mapping 
(\ref{bogolubov  mapping a}) with (\ref{bogolubov  mapping mu nu}),  yielding to 
\begin{align}
& _0\bra{n} N \ket{n}_0= n+ (1+2 n)|\nu|^2\,\, , \qquad  \label{media N su ho nu} 
\\
& _0\bra{n} N^2 \ket{n}_0=n^2+ 2 (3n^2+2n+1)  |\nu | ^2+ 3  |\nu | ^4 (2n^2+2n+1) \,\, ,
\label{medie N N2}
\end{align}
so that
\beq
g_2=\frac{ n(n-1)+ (6n^2+2n+1)  |\nu | ^2+3  |\nu | ^4 (2n^2+2n+1)}{ \left[  n+ (1+2 n)|\nu|^2 \, \right]^2}  \,\, . 
\label{g2 nu}
\eeq
The creation of modes out of a state (e.g. the vacuum) is determined by the square modulus of the Bogolubov coefficient $\nu$.
At early stages of dynamics the process follows a power-law with  scale exponent determined
on account of initial conditions superimposed to $\sigma$ and $\dot{\sigma}$, and of initial values for parametric frequency and its t-differentials. 
Working with Lewis-Riesenfeld invariant number states based on  the minimal uncertainty  conditions (\ref{eq21}), 
one has $|\nu(t)|^2\simeq  \frac{  \sqrt{c} }{4 \omega_0 \sigma_0^2} \dot{\omega}_0^2 t^4+ h.o.t.$
if $\dot{\omega}_0\neq0$ and $|\nu(t)|^2 = \frac{  \sqrt{c} \ddot{\omega}_0^2}{ 36  \omega_0 \sigma_0^2} t^6+h.o.t.$ 
 if $\dot{\omega}_0=0$, 
the approximate form for the leading correction to zero delay correlation function $g_2$ following accordingly via 
$g_2 \simeq 1 -\frac{1}{n} +\frac{2n^2+4n+3}{n^2} \, |\nu|^2$.  
For the problem under consideration in this communication, in particular, the distinction relies on the value for time-translation parameter $t_s$ in (\ref{omega2 gen}) since $\dot{\omega}_0=0$ for  $t_s=0$. 
Once  it is assumed that minimal uncertainty is realized at initial time for elements of the invariant Fock space 
owing to  (\ref{eq21}), the initial variance of the invariant number operator is  vanishing 
 and the initial value for correlation function $g_2$ is the standard  $1-1/n$ (with $n \neq 0$) associated with the harmonic oscillator.

 \subsection{Case 1: $D_e, V_\infty<0$ }
 \label{ss 7 1}
 
\begin{figure}[h!]
\centering
\includegraphics[height=3.8 cm]{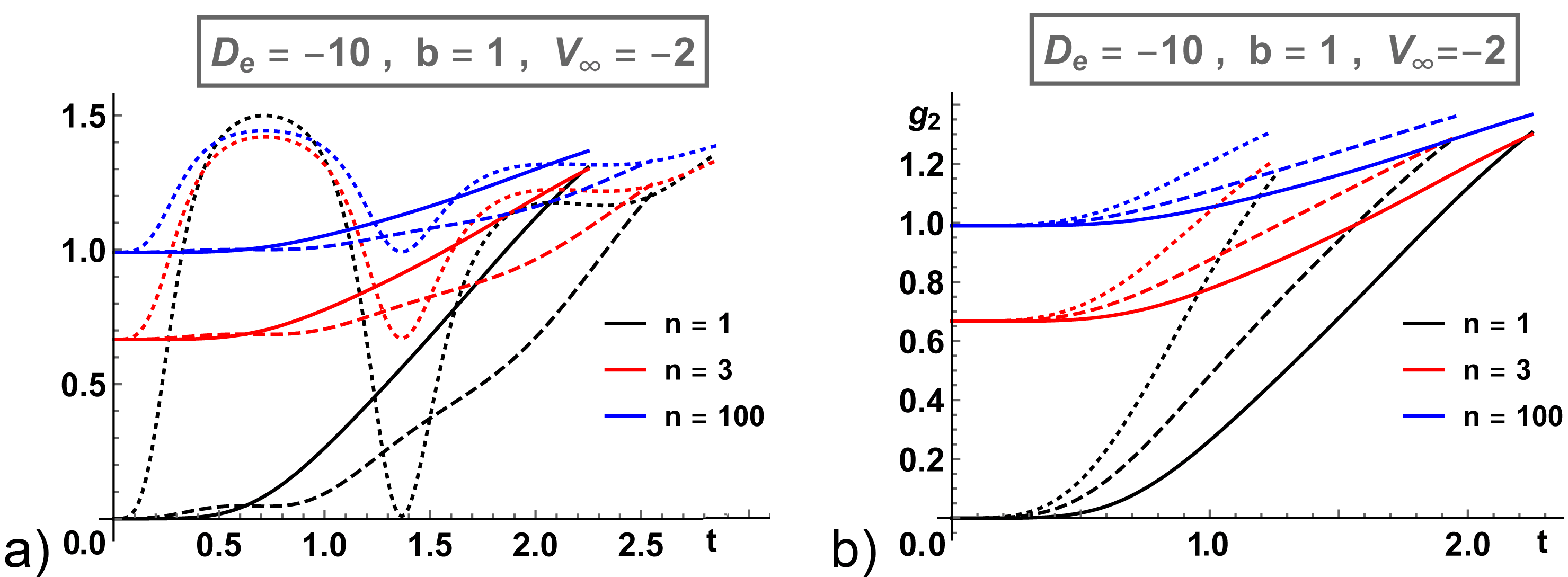} 
\caption{ Second-order correlation functions (\ref{def g2}) evaluated over $\ket{n}_0$ with $n= 1,3,100$,
Eq. (\ref{g2 nu}). 
Left panel: $t_s=-0.6$ (dotted), $-0.3$ (dashed), $0$  (solid);  Right panel: $t_s=0$ (solid), $0.3$ (dashed), $1$ (dotted). When $t_s<0$ substantial deformations of $g_2$ curves originate and  switches  between sub- and super-poissonian regimes are realized. }
\label{figure g2 caso 1}
 \end{figure}

In general, for fixed frequency parameters the curves for $g_2$ assume higher values by increasing the occupation number $n$. Figures \ref{figure g2 caso 1} subsume the essential features displayed when one is concerned with a parametric frequency associated with the case, Eq. (\ref{omega2 gen}) with $D_e,V_\infty<0$. 

If  $\underline{t_s=0}$,  there are no signs of undulations. The system starts to evolve from an adiabatic condition and initial variations of frequency parameter $\omega$ 
do not affect much  the 2-point correlation function. But after a certain time $g_2$ begins to increase its  values almost linearly until reaching its maximum at the final time 
$t_+$, Eq.  (\ref{root tm}). So there is a mechanism acted by the driving parametric frequency that steadily directs  the sub-poissonian initial statistics to a super-poissonian one eventually. Needless  to say, for the occupation number $n=1$ state the process is longer, but when the memory of adiabatic initial condition is lost    the rate of growth 
is noticeably higher than for other number states. Such rate decreases with the increasing of the occupation number, albeit very soon the differences become insignificant. 
At the final time  values for $g_2$ group closely.  Figures \ref{figure g2 caso 1} also  show what happens whenever a shift of the frequency curve is performed by means of  
translations of time variable. For positive shifts $t_s$ the time interval where dynamics takes place is shorter, and it turns out that the growing manifests itself with a greater 
rate but the final maximum value slightly decreases.   It goes without saying that for negative time shift parameter dynamical responses for $g_2$ change. 
For small negative $t_s$ there are marks of fluctuations forming.  For larger negative $t_s$ a well defined bell-like profile characterizes $g_2$, whose maximum develops 
about the time $t=|t_s|$ when the parametric frequency displays its extremum. The top of the bell is in the $g_2>1$ portion of the plane, and hence it is concerned with 
super-poissonian characteristics. A dip follows that drives the statistics to sub-poissonian regime for a short time interval (getting shorter with the increasing of the occupation 
number), unless one is concerned with rather high occupation numbers (with parameters of figure \ref{figure g2 caso 1}.a) and $t_s=-0.6$ one gets $n>422$, for instance).  
A renewed propensity to increase bunching follows guiding back to super-poissonian features, followed by another change of signs for differentials  of minor extend.    

\begin{figure}[h!]
\centering
\includegraphics[height=3.8 cm]{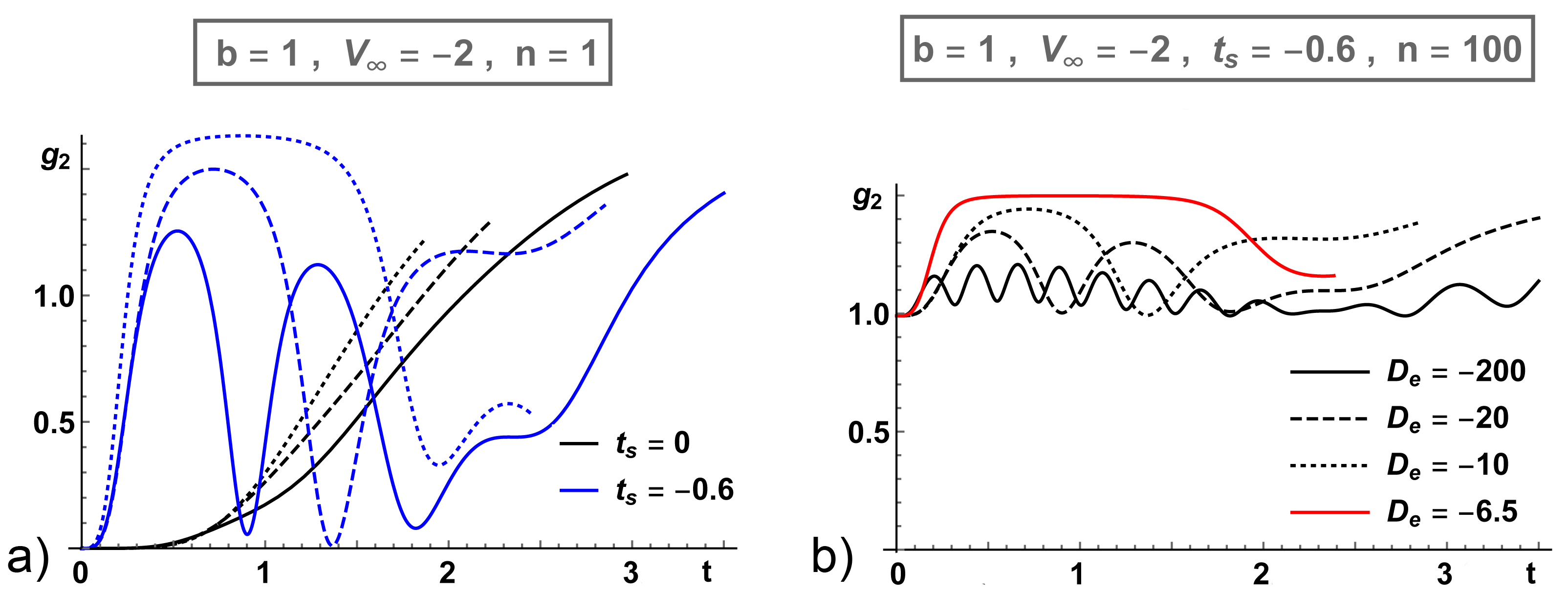} 
\includegraphics[height=3.8 cm]{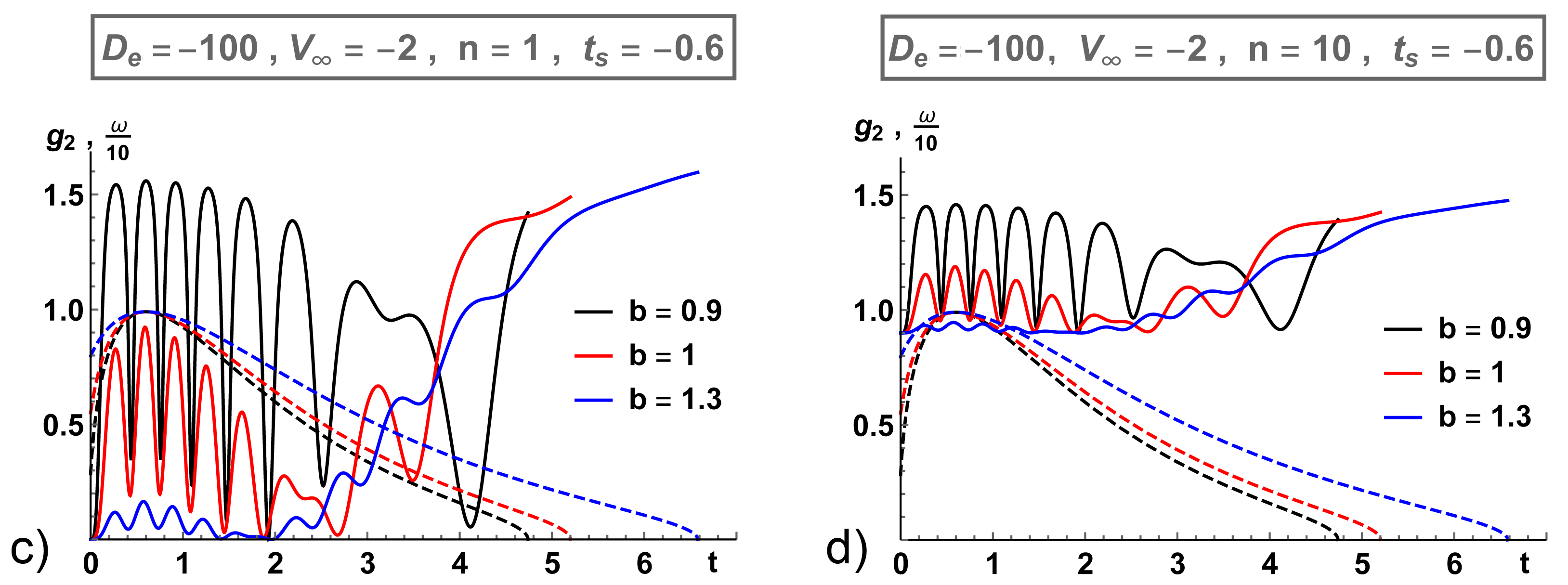} 
\includegraphics[height=3.8 cm]{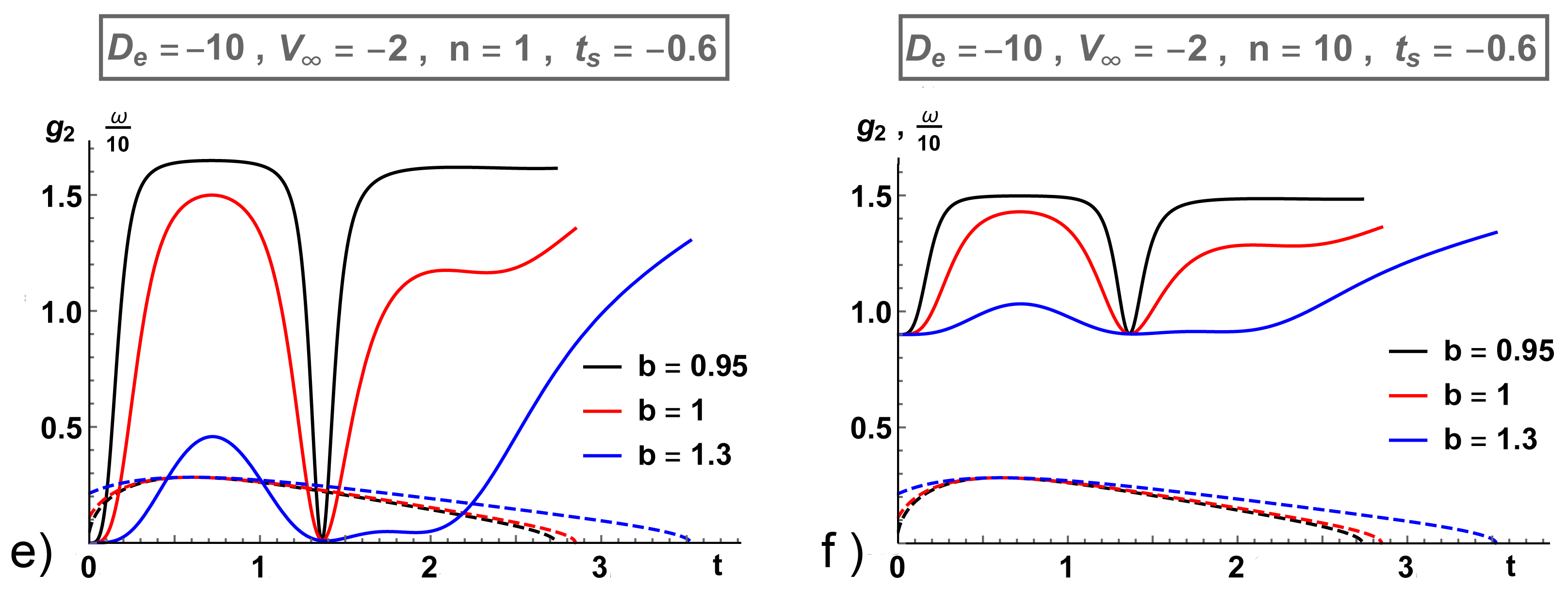} 
\caption{   Case 1: second-order correlation functions evaluated over  harmonic oscillator eigenstates. a) $n=1$ and $D_e=-20, -10, -7$ (solid, dashed, dotted); b) $n=10$. 
By adjusting the frequency parameters, essential changes affect the particle creation mechanism and remould $g_2$, with  more squared shapes or more peaks emerging, or with suppression/amplification of fluctuations. For low quantum numbers, inverted spikes can be seen due to the sudden changes in the dynamics. }
\label{figure g2 caso 1 bis}
 \end{figure}

What happens by varying the intensity of the Morse term when $t_s<0$ is shown in graphics \ref{figure g2 caso 1 bis}. Curves tend to lower or uprise as a consequence 
of changes in  initial condition and time-window for dynamics.  Higher values of $D_e$ result in the enlargement of the bell, and in the replacement of the maximum peak 
with a bigger and bigger plateau. In contrast, lower values of parameter $D_e$ introduce more oscillation of lower magnitude, and hence more changes in the statistics. 
For $t_s>0$, one would see instead that the decrease of $D_e$ merely tends to lowering curves \ref{figure g2 caso 1}.b) and  spreading them in a wider time interval.

 It is now important to see what happens by refining the parameter $b$ which fixes the scale for the time-evolution parameter $t$. 
The operation really has a repercussion on the determination of $\omega_0$,  and  of adiabaticity of  $\omega$
and $\sigma$.  Indeed, we can see that the dynamics of correlation functions $g_2$ contrasts  dramatically  with that  comprehended previously for $b=1$. For higher values of $b$, the adiabaticity of the driving improves: $\omega_0$ approaches the maximum value of $\omega(t)$; in addition, the parametric frequency curve stretches out on a wider time domain, Eq. (\ref{root tm}). The initial prominence previously seen for the correlation function $g_2$ is appreciably dampened, and the successive going up again  is evidently  done much less quickly. 
In contrast, for lower values of $b$, the previously observed initial bell-like shape turns into a more squared profile, with the formation of a second  plateau.  
The role b has in suppression or enhancement of sudden changes is obviously lead to situations based on different choices for the other parameters defining the frequency parameter.   In particular, we have already noted through Figs. \ref{figure g2 caso 1 bis}.a) and \ref{figure g2 caso 1 bis}.b) that lowering $D_e$ introduces more fluctuations   
for the correlation function, though with lessened magnitudes. 
Figures \ref{figure g2 caso 1 bis}.c) and \ref{figure g2 caso 1 bis}.d) present what results if a variation for $b$ is performed in combination 
the assumption of large values for $D_e$. In these figures,  $D_e=-100$, rather compressed and wide fluctuations for $g_2$  are shown that are generated in the first part time evolution for $b=1$, and wider ones,  with higher mean values,  are sustained once $b$ diminishes. In the remaining part of evolution,  that pattern breaks, to eventually conduct (with or without fluctuations, depending on the case)  the function  $g_2$  in the super-poissonian regime. For higher values of $b$,  the initial oscillations are strongly suppressed and values for $g_2$ in the first part of dynamics also are much contained. For the  second part of evolution  a  stretching of the $b=1$ curve over the enlarged  time interval appears.

\subsection{Case 2:  $D_e<0$  and $V_\infty>0$.}
\label{ss 7 2}

Bearing in mind the discussion in Section \ref{ss 5} concerning the amplitude $\sigma $ dynamics,   one is aware beforehand of the incidence of oscillatory features  for the parametric oscillator problem  when evolution is governed by the pumping mechanism of Cases 2 or 3 defined in the infinite time interval $ t\in [0,\infty)$.  
Specifically, translations of the Morse-type curve for $\omega^2$ are expected to give rise to a gradual developing of oscillation in the $g_2$ spectrum. 

\begin{figure}[h!]
\begin{center}
\includegraphics[height=3.8 cm]{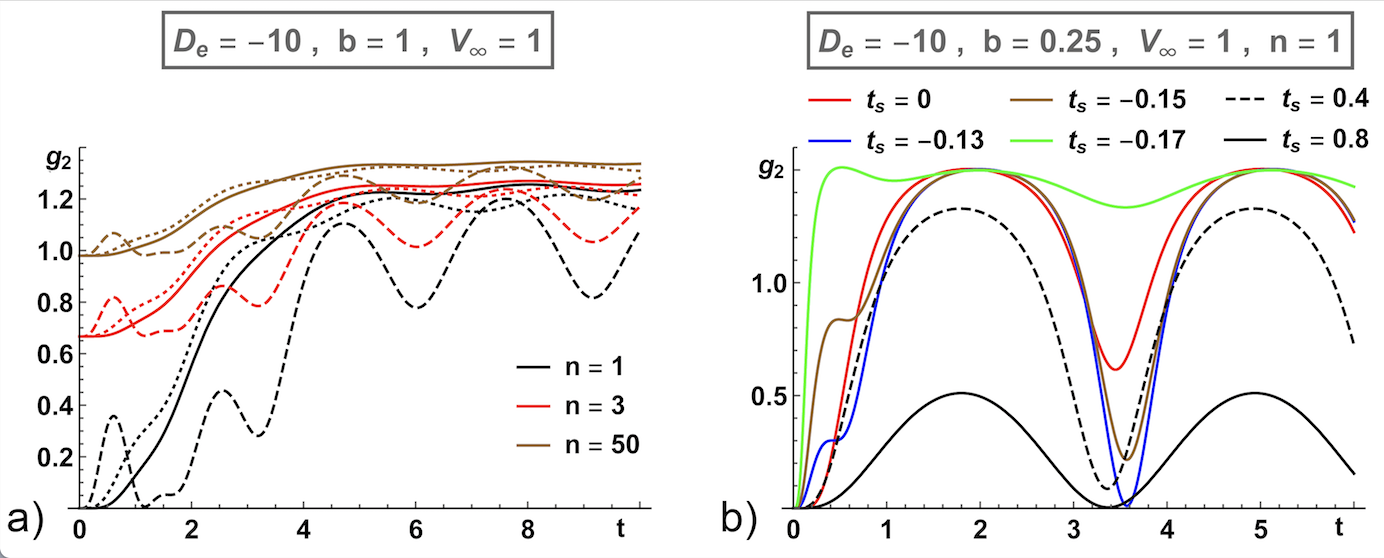}
\includegraphics[height=3.8 cm]{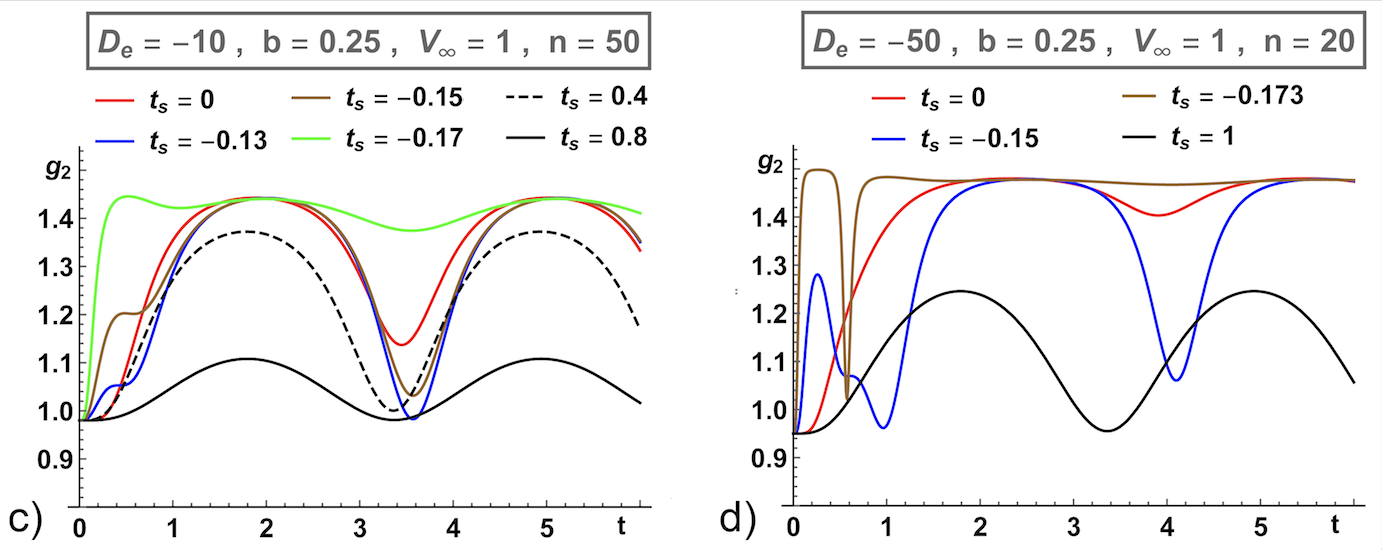}
\end{center}
\caption{Case 2: $g_2$ over number states. 
a)  $t_s= -0.5, 0, 0.5$ (dashed,solid, dotted); b) $n, D_e, V_\infty$ are as in a), but with smaller $b$ and different $t_s$; c)  $n=50$ and $b, D_e, V_\infty$ as in figure b). 
d) $n=20$ and $b, V_\infty$ as in previous two plots.   Figures show oscillation patterns forming after an initial time transient. 
Oscillation may be as enlarged to reach the initial values and to provoke a bouncing of curves along with reduced minima/maxima excursions, see b)-c). For greater values of 
$|D_e|$ an inverted spike is seen to develop for the lowest allowed values of $t_s$. }
\label{g2 caso 2}
\end{figure}

In Figure \ref{g2 caso 2} we have plotted typical curves for the correlation function $g_2$ that elucidate  the various  influences of the shift parameter $t_s$ entering the parametric frequency. 
If $\underline{t_s=0}$,  an initial augment for  $g_2$ that reminds of that seen in case  1 is followed by a sort of asymptotic saturation with extremely  weak signs of fluctuations.  When one is concerned with high occupation numbers, a super-poissonian regime is obviously realized in relatively short time. The major differences one has in the initial magnitude of $g_2$ for the very low occupation numbers vanishes in the course of dynamics, as curves tend asymptotically to attain values in a small domain. Being more attentive  to  the first trait of the curves,
it is seen that the successive minor fluctuations are in fact anticipated by rather unperceivable variations in the rate of the function's growth.
So it is intuitive that the case possesses the characteristics for sustaining large oscillations in the $g_2$ spectra whenever changes are introduced via nontrivial time shifts $t_s\neq0$. 
If $\underline{t_s<0}$, pronounced oscillations indeed arise owing to the major degree of non-adiabaticity of frequency variations to which states are subjected in  their initial dynamics. A regime of regularity in the oscillator  of $g_2$ appears after a transient time. In the first course of dynamics, the frequency increases to reach its maximum, 
and a first local maxima is in fact  exhibited by $g_2$ through a peak above the $t_s=0$ curve. 
After this peak, the $g_2$'s  profile stays below the $t_s=0$ curve, but goes towards it oscillating. When sufficient time is passed, the curve for $g_2$ with $t_s<0$ sustains more regular oscillations, whose maxima get close to the $t_s=0$ curve. The increasing of the occupation number let the  magnitude of shape variations diminish. 
If $\underline{t_s>0}$, obviously departures  from the $t_0$ curves  are realized to a minor extend. By comparison with plots when $t_s=0$, 
a bit higher values can be detected in the initial growth of  $g_2$, and lower ones  later on.

It is now interesting to comprehend what happens by refining the time scale where the exponential terms in the frequency act more importantly.  In view of findings in Subsection \ref{ss 7 1}, we expect  appreciable alterations of the previous big picture. 
Figure \ref{g2 caso 2}.b)  shows an example on the consequences a scaling $b\to b/4$ produces if all the other parameters of figure (\ref{g2 caso 2}) are unaltered. If we take  $n=1$ (the case experiencing the greater excursions of values assumed), we can see that  large oscillations show up even when $t_s=0$, and $\dot{\omega}$ vanishes at initial time. Alternation of sub/super poissonian  changes occurs.  
Increasing positive $t_s$  both the maxima and the magnitude of oscillations progressively reduce, and a damping mechanism acts on the curves. For $t_s<0$ a different process operates that in the end spoils fluctuations. 
For low negative $t_s$  oscillations actually increase, but  later they are inhibited:  when $t_s$  is decreased further, the curve is gradually lifted up to a reference stationary value, reached in very short time. For low negative $t_s$ the minima attained in the course of oscillation  approaches the initial value for $g_2$. 
When $t_s$  is decreased further, curves $g_2$ are bounced back. The initial value for $g_2$ sets a reference bound value also quantum numbers $n$ are greater than $1$, an instance being given in  Figure \ref{g2 caso 2}.c).  
As for counting statistics, the parametric driving  is capable to imprint an essentially super-poissonian behavior, unless very low quantum occupation numbers $n$ and monotonically varying frequency $\omega$ are concerned.

\subsection{Case 3:  $D_e , V_\infty>0$}
\label{ss 7 3}
\begin{figure}[h!]
\centering
\includegraphics[height=3.8 cm]{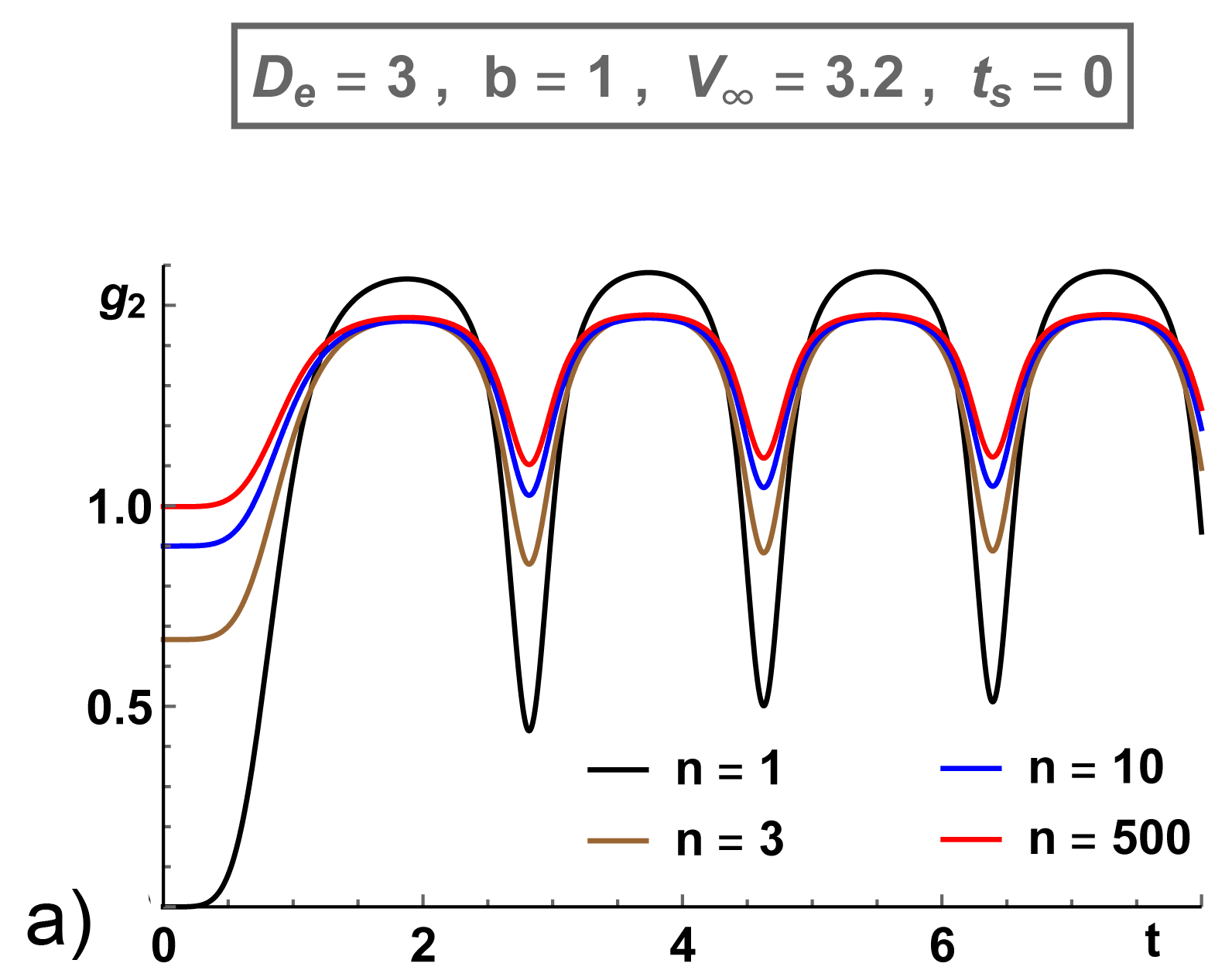}\quad
\includegraphics[height=3.8 cm]{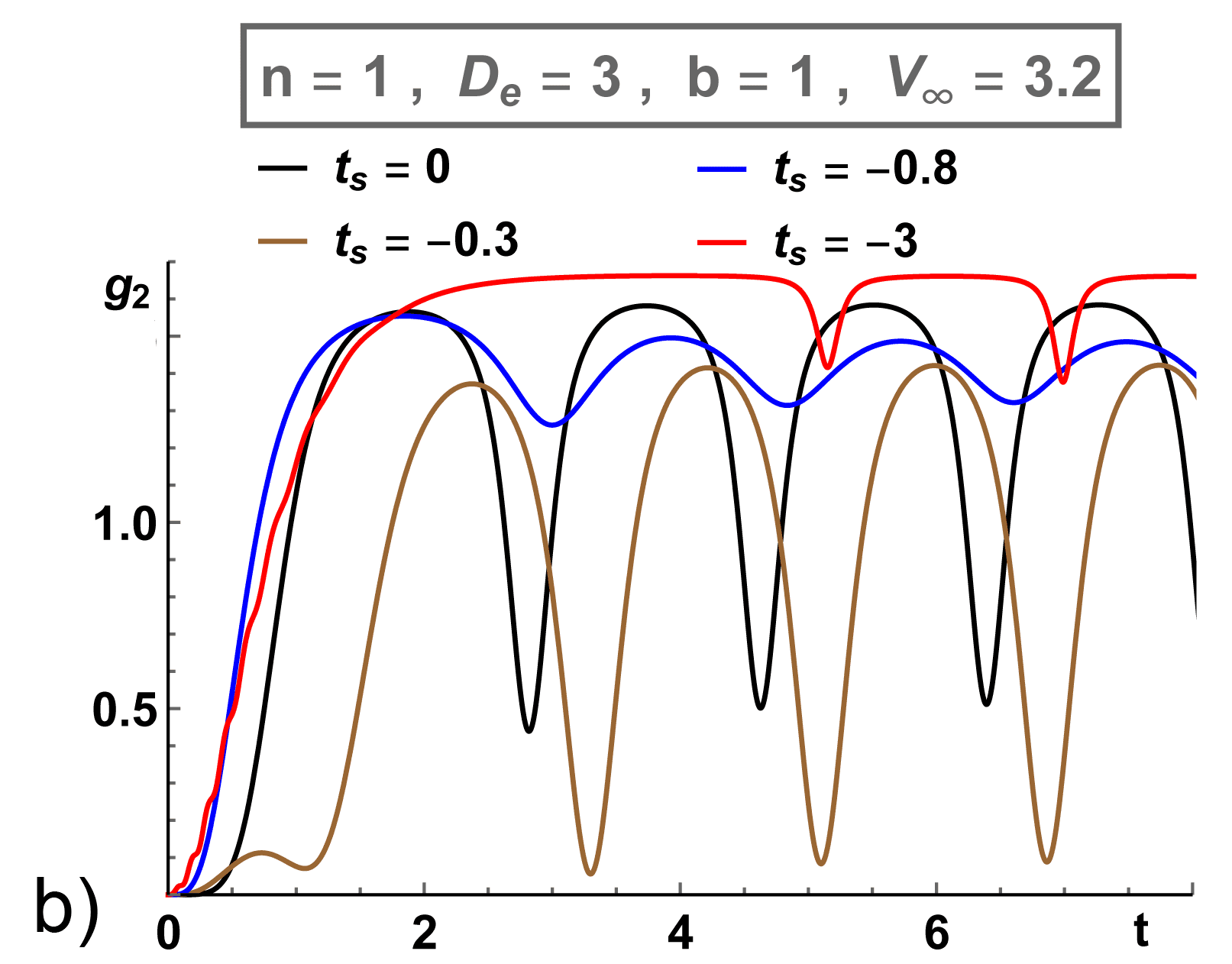}\quad
\includegraphics[height=3.8 cm]{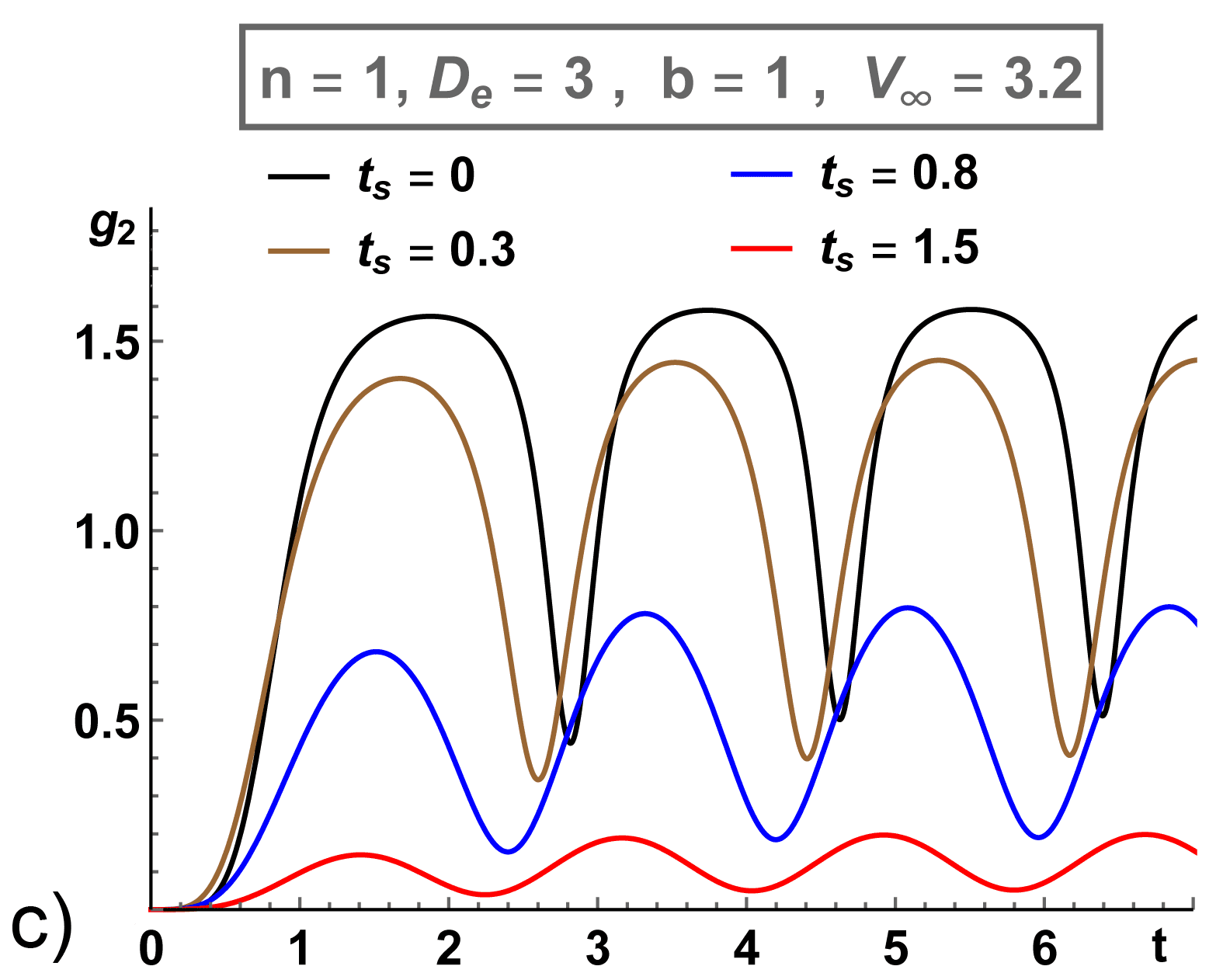}

\caption{Case 3: $g_2$ over number states. a) $t_s=0$ and $n=1,3,10,100$; b) negative $t_s$ and $n=1$; c) positive $t_s$ and $n=1$. 
Inverted spikes are introduced in the oscillating pattern that are smoothed out with the increase of the quantum number $n$ and the tuning of $t_s$.}
\label{fg2 caso 3}
\end{figure}

Plotting the zero time delay correlation function $g_2$ versus time we see the expected generation of fluctuating patterns. Figures \ref{fg2 caso 3} give the gist of variation of occupation number $n$ and time-shifts $t_s$.
The oscillating pattern is flattened down by higher values of positive $t_s$, as upshot of the declined initial variation and non-adiabaticity experienced by the parametric frequency. 
\footnote{
\noindent Both in this case and the previous one of  Subsection \ref{ss 7 2}, Bogolubov coefficients
$
\mu_\infty=\frac{\omega_\infty+\omega_0}{2\sqrt{\omega_\infty\, \omega_0}}$ and $\nu_\infty=\frac{\omega_\infty-\omega_0}{2\sqrt{\omega_\infty\, \omega_0}} $ 
referring to an adiabatic asymptotic limit may be introduced to approximately assess at the lowest order the modification of statistics at large times.}
For small negative values of $t_s$, one whiteness to a right and down translation of the profile.  This mechanism  is accompanied by  the birth of a new initial hump that, once lower values for $t_s$ are assumed, it is lifted up with all the others present about the portion of the $t_s=0$ curve that attains maximal values. Diminishing further the parameter $t_s$,  a saturation character emerges: plateau  form that are interrupted by the small holes generated by the minimum-maximum  excursions traits being suppressed and pushed up.  
 
\subsection{Case 4: $D_e>0$ and $V_\infty<0$}
\label{ss 7 4}
 
 \begin{figure}[h!]
 \centering
\quad \includegraphics[height=3.6 cm]{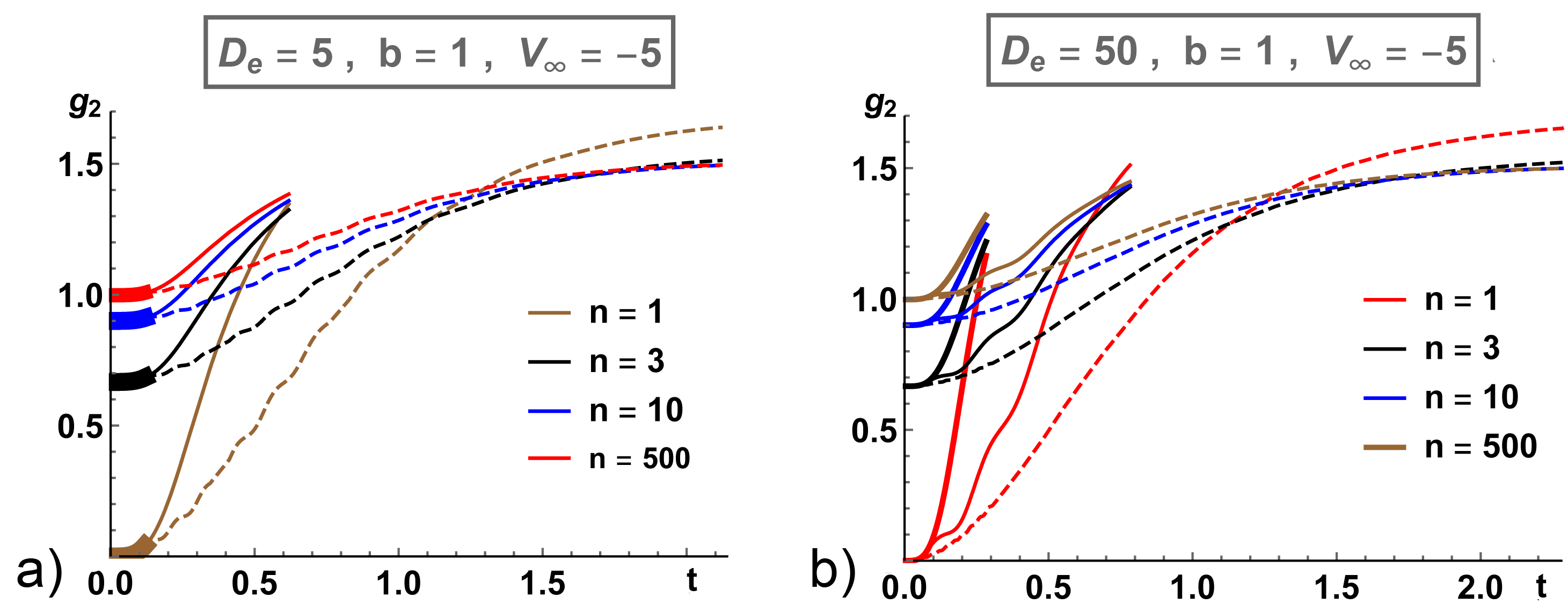}  
\includegraphics[height=3.6 cm]{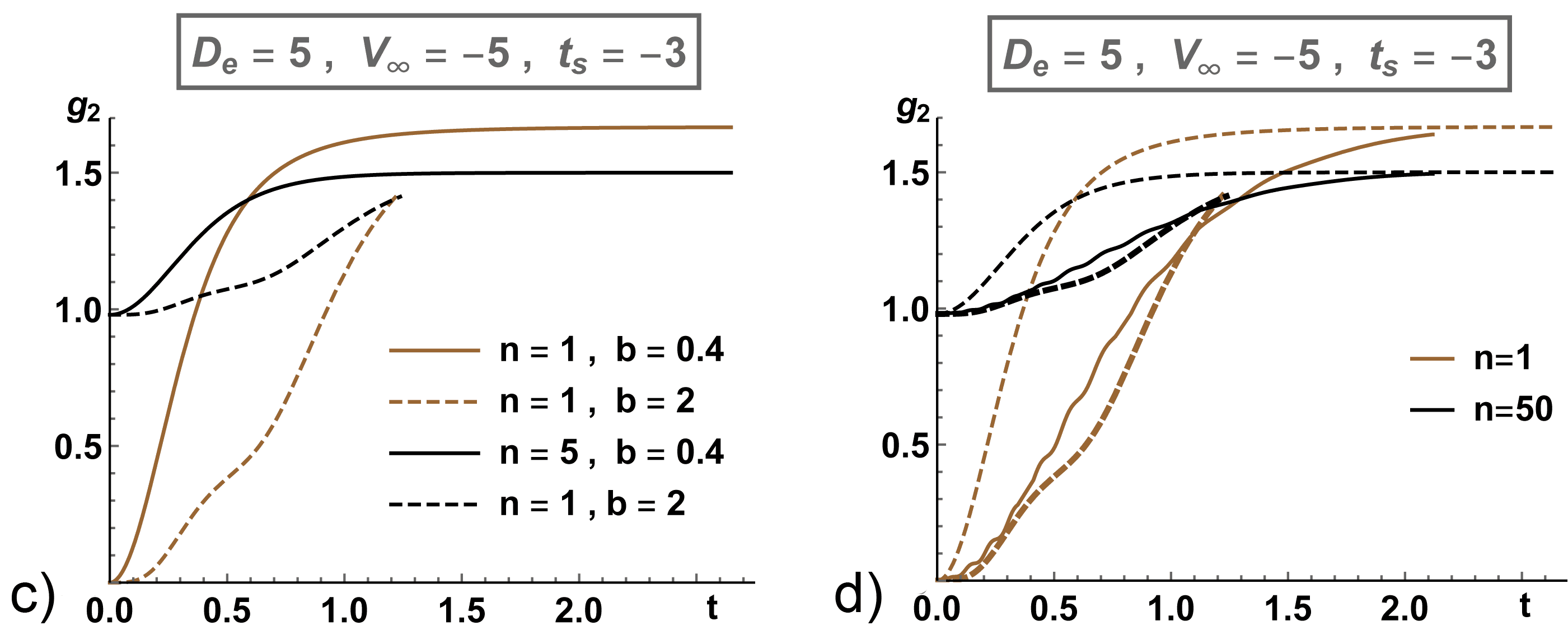} 
\caption{Examples of $g_2$ over number states in Case 4 for different parametric frequencies.  
a)-b) $b=1$, $V_\infty=-5$  and  $t_s=-1,-1.5,-3$ (solid thick, solid, dashed); $D_e=5,\,50$.
c) $D_e=5$,  $V_\infty=-5$, $t_s=-3$, $b=0.4, 2$ and $n=1,\,5$. d) $D_e=5$,  $V_\infty=-5$, $t_s=-3$, $b=0.4, 1, 2$ (dashed, continuous, dashed thick) and $n=1,\,50$. 
 $g_2$ is subjected to a fast growing behavior turning into a plateau if the parametric driving is active for sufficiently long time.}
\label{fg2 caso 4}
\end{figure}

The driving mechanism  exerted  by the parametric frequency now runs for a short  time interval (in $b$ units) and 
selecting the most rapidly varying portion of the Morse curve. 
Examples of functions $g_2$ are provided in Figure \ref{fg2 caso 4}, showing how  they move into the $g_2>1$ region if the system evolves for a sufficiently large time.
For intermediate values of negative parameter $t_s$, the case it is seen to share some similarities with Case 1 with $t_s\geq 0$,  Fig. \ref{figure g2 caso 1}.b).  
The duration of time interval where the driving mechanism is active   does not get through $t_+$, Eq. (\ref{root tm}),   creasing limitedly for higher values of $D_e$ and $t_s$. 
Effects of such an enlarging the time window where the dynamics takes place are visible with possible  vague signs of prior undulations, and with an additional trait
suggesting that it would likely progress to a saturation. It is the acting on the value of time-scale parameter $b$ that allows the saturation to take a more concrete form, 
by simultaneously  smoothing  the modest undulations. As we have noticed in Subsection \ref{ss 5 4}, most influential alterations for the frequency parameter
are indeed those of parameter b, whose reduction gives a notable boost to values for 
$\omega_0$ and the rate of changes for $\omega$. These  then imply a strong suppression of the initial value $\sigma_0$ that is assumed via (\ref{eq21}) for the solution $\sigma$ to the Ermakov equation, and an enhancement of the variation  of $\sigma$ that can be observed at later  times compared to its initial value. 
This makes a difference because the Bogolubov coefficient $\nu$ is quite sensitive to quantitative changes experienced by the function $\sigma / \sigma_0$. 
When $\sigma / \sigma_0$  deviates from values that are close the unity,  a growth for $|\nu|$ is  dictated that  easily can be sufficient to determine the $|\nu|^4$ terms dominating $g_2$. Under these circumstances, at a certain point the mode creation mechanism can be therefore strong enough to let $g_2$
approach a quality of stationarity.  
 
\section{Delayed second order correlation functions}
\label{ss 8}

To complete the understanding of statistical aspects we earned so far through the zero-delay correlation functions, 
it is helpful to  considering also the two-time correlation function over states $\ket{n}_0$, i.e. the function
\beq
g_2(\tau)=\frac{ _0\bra{n} a^\dagger(t) \,a^\dagger(t+\tau) \,a(t+\tau) \,a(t) \ket{n}_0 }{ _0\bra{n} a^\dagger(t) a(t) \ket{n}_0 } \,\, .
\eeq
This enables one to get an insight into the fulfillment of  anti-bunching/bunching  requisites and the possibility to 
measuring modes in a time-delayed fashion with higher probability than being measured at the same time. 
We will limit ourselves to the cases where $D_e$ and $V_\infty$ are both either positive or negative (Cases 1 and 3) because they allow to epitomize the 
basic features  displayed also for different choices of the  driving frequency parameters.  Yet, we shall focus on correlations over  the lowest excited 
 states $\ket{n}_0$ as  they present wider   domains where anti-bunching is accomplished.

\begin{figure}[h!]
\begin{center}
\includegraphics[scale=0.290]{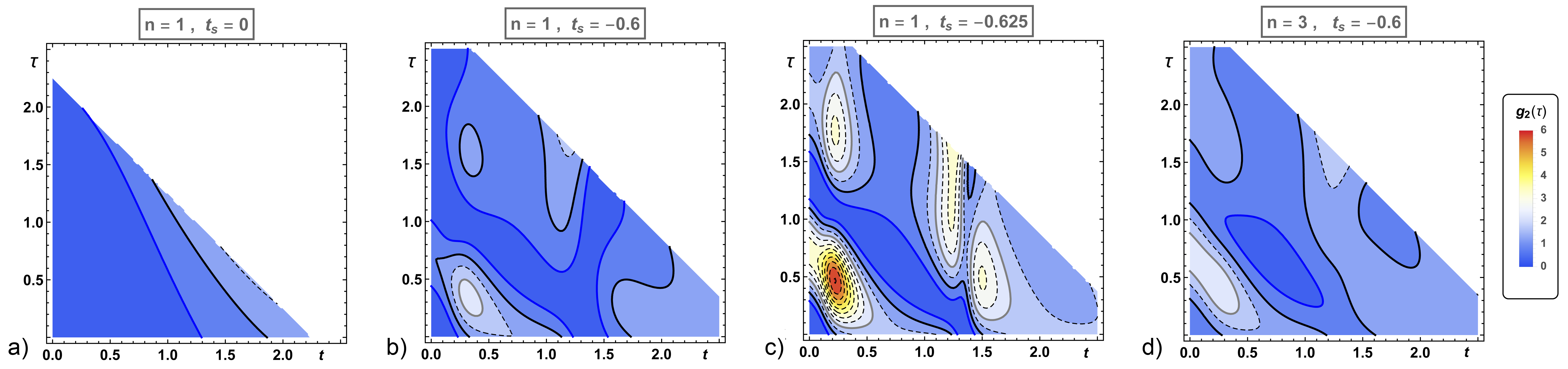}
\end{center}
\caption{Case 1: parametric frequency (\ref{omega2 gen}) with $D_e,V_\infty<0$. 
Level plots for the delayed correlation function $g_2(\tau)$ when   $D_e=-10$ and $V_\infty=-2$. 
 Thick solid curves refer to $g_2(\tau)=1/2$ (blue), $g_2(\tau)=1$ (black), $g_2(\tau)=2$ (gray).
 a)-c) Dynamics underlying the reduction of anti-bunching and the surfacing of strong bunching when $n=1$ and 
the parametric frequency initial shape is modified by assuming $t_s=0,-0.6,-0.625$. d) $n=3$ and $t_s=-0.6$; by comparison with figure b)  
the effect of changing the reference state is noted.} 
\label{fig g2tau 1}
  \end{figure}

Plots in Figure \ref{fig g2tau 1} provide instances of  the bunching and anti-bunching mechanisms possibly realized by considering different  time-translation parameters \underline{$t_s\leq 0$} for  the parametric frequency (\ref{omega2 gen}) with $D_e,V_\infty<0$. 
Having a finite time interval over which the dynamics takes place clearly introduces a finite domain for the delay variable $\tau$.
If $t_s=0$ and the very low occupation numbers $n$ are concerned, there is a mostly anti-bunched behavior that is displayed by varying the time $\tau$ determining the second counting. For $n=1$ a relatively vast area denoting strong anti-bunching ($g_2(\tau)<1/2$) obviously arises, a strong anti-bunching being sustained for all the allowed $\tau$ for lower $t$'s -Fig. \ref{fig g2tau 1}.a).  Diminishing $t_s$, the moderate extension of the time domain allowed for evolution causes an enlargement of the region in the $(t,\tau)$ plane that identifies anti-bunched responses; this also holds for the strongly anti-bunched domain appearing for $n=1$. At this stage, contour plots for $g_2(\tau)$ thus qualitatively resemble those obtained for  $t_s=0$. By diminishing further the values of $t_s$ so to have less adiabatic initial changes for the parametric frequency and more pronounced fluctuation effects for physical quantities, the picture for second-order correlation functions modifies: islands  
reflecting bunching phenomena are introduced, possibly tending to connect each to the other while introducing higher values for $g_2(\tau)$. Remark that
the condition  $t_s<0$ (which implies a change of sign for $\dot{\omega}$ at early times) does not exclude the possibility to support the detection of strong anti-bunching $g_2(\tau)<1/2$ when $n=1$, even though an overall inclination to build bunched super-poissonian responses may be displayed. 
Figure  \ref{fig g2tau 1}.b) shows for instance that for $n=1$ a strong anti-bunched path can be identified  in plane $(t,\tau)$ that originates when $t_s$ is about $1.4$ and $\tau$ is varied in its whole allowed domain. 
For the closely successive values of the quantum number $n$, the behavior of correlations $g_2(\tau)$ is affected accordingly, see Figure \ref{fig g2tau 1}.d)  
determined for $n=3$ and with the same frequency parameters $D_e, b, V_\infty$ and $t_s$ of Figure \ref{fig g2tau 1}).b).
The zero-delayed second order correlation function does not present  strongly sub-poissonian traits as in the $n=1$ case, but delayed correlation can attain low values. 
In  Fig. \ref{fig g2tau 1}.d)  it is seen that strong bunching or strong anti-bunching can be detected.  \\
  \begin{figure}[h!]
  \begin{center}
  \includegraphics[scale=0.0685]{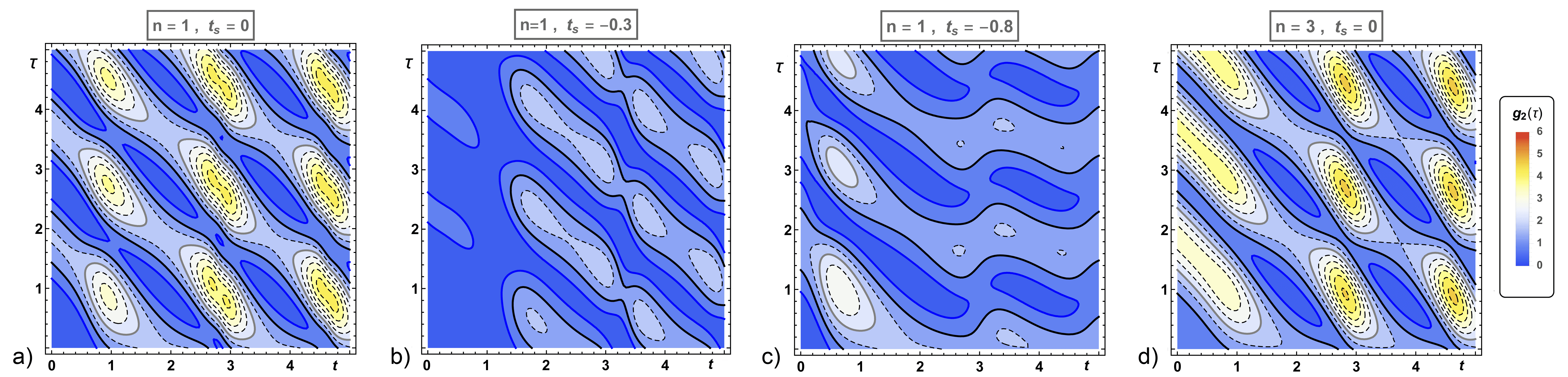}
    \end{center}
\caption{Case 3: parametric frequency (\ref{omega2 gen}) with $D_e,V_\infty>0$.  Level plots for the Glauber function $g_2(\tau)$ over number states $\ket{n}_0$ when $D_e=3$ and $V_\infty=3.2$. Thick solid curves refer to $g_2(\tau)=1/2$ (blue), $g_2(\tau)=1$ (black), $g_2(\tau)=2$ (gray).
a)-c) Revival and collapse of bunching and antibunching  when $n=1$ and 
the parametric frequency initial shape is modified by assuming $t_s=0,-0.3,-0.8$. d) $n=3$ and $t_s=-0.8$; by comparison with figure c)  
the effect of changing the reference state is noted.
} 
\label{fig g2tau 3}
  \end{figure}
When $D_e, V_\infty>0$ in (\ref{omega2 gen})  there is no upper time limits for the dynamics and regular oscillating patterns generate after the initial transient. These circumstances are reflected in functions $g_2(\tau)$, and the tendency  to develop repeating patterns is recognized. 
Figures \ref{fig g2tau 3}.a)-c) show, in particular,  examples referring to the two-time correlations over the states $\ket{1}_0$. 
The alternation of domains associated with strong manifestations of anti-bunching and bunching is visible in Figure  \ref{fig g2tau 3}.a), concerned with a case for which $t_s=0$. 
Upon decreasing the parameter $t_s$ patterns modify, a suppression on responses takes place and larger islands associated with anti-bunching show up for $\tau>0$, Fig. 
\ref{fig g2tau 3}.b).  
By further decreasing $t_s$  the foreseen  reinforcement for the mechanism of bunching follow, Fig. \ref{fig g2tau 3}.c),  
 with higher values for $g_2(\tau)$ attained by increasing the quantum number $n$, Fig. \ref{fig g2tau 3}.d).

\section{Conclusions}
\label{ss 9}

 We have discussed the dynamics of one-degree of freedom Hamiltonian of the parametric oscillator type with a time-dependent frequency based on a Morse potential shape, up to possible translation and sign reversion. 
This provided us an analytical  framework to examine both cases where there is a monotonic increasing or decreasing energy pumping mechanism, and cases where there is an alternance of them  realized   continuously. Depending on choices of the parameters, 
the pumping mechanism can be active unlimited or  cease  after a certain time interval.

For time-dependent quadratic systems, solution to the Sch\"odinger equation and quantum dynamics are manifestly based on classical dynamics
 as the centroid and the spreads of a wave-packet follow exactly the evolution of a classical particle \cite{ littlejohn,andrews,  dieter book, manko}. 
We thus detailed  (Appendix \ref{appendice}) the derivation of solution to classical equation of motion for the parametric oscillator with the time-dependent frequency we set out and (Section \ref{ss 5})  analyzed the features of their amplitudes $\sigma$,  whose variations are governed by the Ermakov nonlinear differential equation and direct all the dynamical consequences.  Enhancement and suppression of their oscillations by virtue of frequency parameter shape changes have been thoroughly inspected,
with plots summarising the behavior of functions $\sigma$ for different values of the parameters involved and the initial conditions we adopted for the Ermakov differential problem (\ref{eq Ermakov}). Figures show, for instance, the manner Sakharov-type regular oscillatory patterns in systems' responses do form when the parametric frequency modifies to eventually approach a quality of stationarity.  They also illustrate how, when initial conditions and the changes for the parametric driving are 
such that a big (non-adiabatic) effort is needed for the system to adapt itself to the dynamical circumstances,
 larger oscillation develop about the leading order curve from the WKBJ expansion method. 
 Progressive enlargement of fluctuations for non-adiabatic changes of amplitudes $\sigma$  adapt closer and closer to 
 (the absolute value of) the complete non-stationary mode solutions to the parametric oscillator equation;  accordingly, the phases $\theta(t)$ of classical 
 modes develop by alternating sudden continuous jumps  to nearly constant traits, a behaviour inherited at the quantum level by the geometrical phases (\ref{eq7}).

  We later argued on  how quantum dynamics mirrors the squeezing mechanism implied at the classical level on the Ermakov invariant orbits on phase-space,
and in this respect we presented plots related to the Heisenberg uncertainty relation on number-type  and  coherent-type states, 
including  some examples showing that these uncertainties can act on a scale greater than $\hbar$.
Finally, to refine the insight into the statistical contents implied, 
  the generation during the course of evolution of nonclassical features for  number-type  states has been discussed by investigating deviation from Poisson statistics of two-point correlation function with zero time delay owing to the squeezing dynamics taking place. Delayed two-point correlations have also been considered to attest
 the phenomena of time-depending bunching and anti-bunching showing up.   
 Results obtained clearly show trends of collapse and revival phenomena whose intensity is conditional on the dynamical facets that have been highlighted for the solution to the Ermakov equation. 
  
 Studies like the one we performed in this communication find a natural arena of application,  e.g., within the context of the electromagnetic wave propagation in 
time-dependent media \cite{nerukh,zhang,lin,pedrosa2011, kalluri}. The analysis 
can be then continued by taking into account more  complicated states or by implementation of quasi-probability tools. 
In particular, linear combination of states can be considered to analyze  the competition between interference effects and parametric frequency variations both on sufficiently long-time behavior and at early stage of the evolution (where they are more consistent), and the consequent effects on correlations. 
The sensitivity of  expectation values to initial conditions other than (\ref{eq21}) can also be checked.
Finally, it definitively appears of interest also to proceed with multi-mode generalization  and to examine dynamics of entanglement. 
We plan to address these issues in the future.

\appendix
	\section{Parametric oscillator equation with Morse-type frequencies: closed form solutions and their Wronskians}
	\label{appendice}
	\setcounter{section}{1}
		\setcounter{equation}{0}
\renewcommand\theequation{A.\arabic{equation}}

In this Appendix,  the derivation of closed form solution to the parametric oscillator equation with frequency (\ref{omega2 gen}) with $V_\infty\neq0$
will be discussed. 
We will perform a step-by-step analysis  drawing special attention on factors affecting the way in which independent solutions for the equation are permitted. 
After that we compute their Wronskians, whose  role  is of central importance as well in explicitly determining  solutions to the Ermakov equation via formulae 
 (\ref{eq30})-(\ref{eq31}). We will provide the analytical treatment for the derivation of independent solutions when $D_e, V_\infty<0$ as the other cases follow 
 by proper algebraic adaptments.

\subsection{Case 1: $D_e, V_\infty<0$ }
\label{ss 4 1}

Let us consider a square parametric frequency defined through  an inverse Morse potential with a negative nonvanishing asymptotic value,
Equation (\ref{omega2 gen})  with $D_e, V_\infty<0$. With a change of variable 
\beq
\xi=2\,d\,b\,e^{-\frac{t+t_s}{b}},
\label{def xi}
\eeq
the homogeneous second order linear differential equation (\ref{eq27})
 reads
\beq
\frac{d^2x(\xi)}{d\xi^2}
+\frac{1}{\xi}\,\frac{dx(\xi)}{d\xi}
+\left( \frac{d\,b}{\xi}-\frac{1}{4} +\frac{v^2 \,b^2 }{\xi^2}\right)x(\xi)=0\,
\label{eq45}
\eeq
where the positive parameters $v=\sqrt{- V_\infty}$ and $d=\sqrt{-D_e}$ have been defined for convenience.
If we let $x(\xi)$ to be
\beq
x(\xi)=e^{-\xi/2}\xi^{b\,v}z(\xi)\,,
\label{eq46}
\eeq
the differential equation (\ref{eq45}) becomes, upon dividing by $e^{-\xi/2}\xi^{b\,v}$, the confluent hypergeometric equation
\beq
\xi \, \frac{d^2 z(\xi)}{d\xi^2}+\left(\beta-\xi\right)\,  \frac{d z(\xi)}{d\xi}-\,\alpha\,z(\xi)=0
\label{eq47}
\eeq
with parameters 
$ \alpha=\frac{1}{2}+b(v-d)$ 
and $\beta=1+2\,b\,v$.
A solution to the confluent hypergeometric equation is  the Kummer's function  of the first kind $_1F_1\left[\alpha;\beta;\xi \right]$ defined as
\beq
_1F_1\left[\alpha;\beta;\xi \right]=\sum_{n=0}^\infty\frac{(\alpha)_n\,\xi^n}{(\beta)_n}\,,
\label{eq49}
\eeq
where $(\alpha)_n=\alpha(\alpha+1)...(\alpha+n-1)$ and $(\alpha)_0=1$.
Note that the Kummer's function $_1F_1\left[\alpha;-n;\xi \right]$ is not defined for $n$ non-negative integer, except for the case $\alpha=-m$ with  $m$ non-negative integer and $m<n$. In these cases, other representations of the solutions to (\ref{eq45}) should be used (for more details here, 
see Refs. \cite{abramowitz}-\cite{zwillinger}).\\
It is easy to show that  $\xi^{1-\beta}\,_1F_1\left[1+\alpha-\beta;2-\beta;\xi \right]$ is another solution to (\ref{eq45}). In fact, by letting $z(\xi)$ be
\beq
z(\xi)=\xi^{1-\beta}\,v(\xi)\,,
\label{eq50}
\eeq
upon dividing by $\xi^{1-\beta}$, equation (\ref{eq45}) becomes:
$\xi\,v(\xi)+\left(2-\beta-\xi\right)\dot{v}(\xi)-\,(1+\alpha-\beta)\,v(\xi)=0$.
We can use a combination of the above solutions to define the Tricomi confluent hypergeometric function $U\left[\alpha;\beta;\xi \right]$ \cite{abramowitz},
\begin{align}
U[\alpha,\beta,\xi]=\frac{\pi}{\sin(\pi\,\beta)}\left( \frac{_1F_1\left[\alpha;\beta;\xi \right]}{\Gamma[\beta]\,\Gamma[1+\alpha-\beta]}-
 \xi^{1-\beta}\,\frac{_1F_1\left[\alpha+1-\beta;2-\beta;\xi \right]}{\Gamma[\alpha]\,\Gamma[2-\beta]} \right).
\label{eq52}
\end{align}
The advantage of the function $U[\alpha,\beta,\xi]$ is that, even if it does not exist for $\beta$ integer, it can be extended analytically for all values of $\beta$.
For most combinations of real or complex $\alpha$ and $\beta$, the two solutions (\ref{eq49}) and (\ref{eq52}) are independent.  However, if  $\alpha$ is a non-positive integer, then  (\ref{eq49}) and (\ref{eq52})  are proportional. This is due to the relation \cite{abramowitz,zwillinger},
\beq
U[-n;\alpha;x]=\frac{(\alpha-1+n)!}{(\alpha-1)!} (-1)^n\,F[-n;\alpha;x] \,\ ,
\label{eq53}
\eeq
(with $n=0,1,2,...$). In these cases we should use $_1F_1\left[\alpha;\beta;\xi \right]$ and $\xi^{1-\beta}\,_1F_1\left[1+\alpha-\beta;2-\beta;\xi \right]$ as independent solutions, when they exist. 

In the light of all above considerations, we can finally write the independent  solutions to Equation (\ref{eq45}) as
\beqa
x_1(\xi)=e^{-\xi/2}\xi^{b\,v}\,U\left[\frac{1}{2}+b(v-d);1+2\,b\,v,\xi \right]\,,
\qquad 
x_2(\xi)=e^{-\xi/2}\xi^{b\,v}\,_1F_1\left[\frac{1}{2}+b(v-d);1+2\,b\,v;\xi \right]\,,
\label{eq54}
\eeqa
that in terms of the time variable $t$  becomes 
\beqa
x_1(t)&=&e^{-d\,b\,e^{-(t+t_s)/b}}\,e^{-t\,v}\,\, U\left[\frac{1}{2}+b(v-d);1+2\,b\,v,2\,d\,b\,e^{-(t+t_s)/b} \right]\,,\nonumber \\
x_2(t)&=&e^{-d\,b\,e^{-(t+t_s)/b}}\,e^{-t\,v}\,\,  _1F_1\left[\frac{1}{2}+b(v-d);1+2\,b\,v;2\,d\,b\,e^{-(t+t_s)/b}\right]\,,
\label{eqA14}
\eeqa
for $\frac{1}{2}+b(v-d) \neq -m$, with $m$ non negative integer, and 
\beqa
\tilde{x}_1(\xi) = e^{-\xi/2}\,\xi^{-b\,v}\,_1F_1\left[\frac{1}{2}-b(v+d);1-2\,b\,v;\xi \right]\,,
\qquad 
\tilde{x}_2(\xi)=e^{-\xi/2}\,\xi^{b\,v}\,_1F_1\left[\frac{1}{2}+b(v-d);1+2\,b\,v;\xi \right]\,, 
\label{eq56}
\eeqa
that in terms of the time variable $t$ becomes
\beqa
\tilde{x}_{1,2}(t)&=&e^{-d\,b\,e^{-(t+t_s)/b}}\,e^{\pm t\,v}\,\,  _1F_1\left[\frac{1}{2} \mp b(v \pm d);1 \mp 2\,b\,v;2\,d\,b\,e^{-(t+t_s)/b} \right]\,,
\label{eqA16}
\eeqa
for $\frac{1}{2}+b(v-d) = -m$, with $m$ non negative integer (we have omitted the multiplying factors 
$e^{\pm t_s\,v}\,(2\,b\,d)^{\mp b\,v}$ in    $x_{1,2}$ and $ \tilde{x}_{1,2}$  that arise after the change of variable $\xi \to t$
). 

The Wronskian of the two solutions (\ref{eq54}) can be easily computed by resorting to the known forms for expansions pertaining small arguments  
$\xi$ of special functions involved. So we can take
\beq
W(x_1(\xi),x_2(\xi))=\frac{\cos[\pi\,b\,(d\,-v)]\,\, \Gamma(\frac{1}{2}+b(d\,+\,v))}{\pi\,\xi}.
\label{eq58}
\eeq
  The Wronskian of solutions (\ref{eqA14})  in the independent variable $t$, is related to the Wronskian in the independent variable $\xi$ by means of the relation 
$W(x_1(t),x_2(t))=-b^{-1} \,\xi\, W(x_1(\xi),x_2(\xi))$, giving explicitly
\beq
W(x_1(t),x_2(t))=-\frac{\cos[\pi\,b\,(d\,-\,v)]\,\,\Gamma[\frac{1}{2}+b(d\,+\,v)]}{\pi\,b}.
\label{eq60}
\eeq
Analogously, the Wronskian of the two solutions (\ref{eq56}) is found in the form
$W(\tilde{x}_1(\xi),\tilde{x}_2(\xi))=-2\,b\,v/\xi$,
and the Wronskian of (\ref{eqA16}) in the variable $t$, simply related to $W(\tilde{x}_1(\xi),\tilde{x}_2(\xi))$
 via $W(\tilde{x}_1(t),\tilde{x}_2(t))=-b^{-1} \,\xi\, W(x_1(\xi),x_2(\xi))$, explicitly becomes $W(x_1(t),x_2(t))=2\,v$.

This completes our derivation of the ingredients for expressing in closed form the solutions to the Ermakov equation with the parametric frequency 
Equation (\ref{omega2 gen})  with $D_e<0$ and $V_\infty < 0$.
Note that when  $D_e<0$ and $V_\infty=0$,  two independent solutions to (\ref{eq27}) can only be obtained for $b\,d \neq m+\frac{1}{2}$, with $m$ non negative integer \cite{abramowitz}. Their analytical expression, together with the expression for their Wronskian, is simply given by substituting $v=0$ into Eqs. (\ref{eqA14}) and (\ref{eq60}). In particular, for $b\,d=\frac{1}{2}+m,\,\,\,(m=0,1,2,\dots)$ 
all the solutions to the confluent hypergeometric equation (\ref{eq47})  are proportional to the Laguerre polynomials $L_m(\xi)$  \cite{abramowitz}.

\subsection{Cases  2, 3 and 4}
\label{ss 4 2}

\noindent {\bf Case 2:  $D_e<0$, $V_\infty>0$.} 
The treatment of Equation  (\ref{eq27}) in the present case follows a route similar to that of Case 1 in Subsection \ref{ss 4 1}, and solutions in the $\xi$ variable (\ref{def xi}) will be obtained after performing the replacement $V_\infty\to -V_\infty$, i.e.  $v \rightarrow i\,v$  into (\ref{eq45}). Since $x_{1}(\xi)$ and $x_2(\xi)$ 
are now complex functions, for the evaluation of $\sigma$ we use the functions $y_1(\xi)=x_1(\xi)+x_1^*(\xi)$ and $y_2(\xi)=x_2(\xi)+x^*_2(\xi)$.
In this way the Wronskian becomes, in the original independent variable t, 
\beq
W(y_1(t),y_2(t))= - \frac{2}{\pi\,b}  
\left\{\cos[\pi\,b\,(d\,-i\,v)]\,  \Gamma\left[\frac{1}{2}+b(d\,+\,i\,v)\right]+ \cos[\pi\,b\,(d\,+i\,v)] \, \,\Gamma\left[\frac{1}{2}+b(d\,-\,i\,v) \right] \right\}.
\label{eq65}
\eeq
\noindent {\bf Case 3:  $D_e, \, V_\infty>0$.}
The solution to  (\ref{eq27}) in the $\xi$ variable will be obtained with  the replacement $\xi \rightarrow i\,\xi$, $v \rightarrow i\,v$ and  $d \rightarrow i\,d$ into formulae of Subsection \ref{ss 4 1}, see e.g.  Eqs. (\ref{def xi}) -(\ref{eq45}). 
Since $x_1(\xi)$ and $x_2(\xi)$ are once again complex functions, for the evaluation of $\sigma$  we proceed as in Case 2. The final expression that we get for the Wronskian is 
\beq
W(y_1(t),y_2(t))= -\frac{2}{\pi\,b}  \left\{ \cos[\,b\,\pi(d-v)]\,\Gamma\left[\frac{1}{2}-i\,b\,(d+v)\right] + \cosh[b\,\pi\,d] \cosh[b\,\pi\,v] \, \Gamma\left[\frac{1}{2}+i\,b(d+v)\right]\right\} \,.
\label{eq66}
\eeq

\noindent {\bf Case 4:  $D_e>0$,  $V_\infty<0$.}
Replacements $\xi \rightarrow i\,\xi$ and  $d \rightarrow i\,d$ are performed into formulae of Sub. \ref{ss 4 1} from (\ref{def xi}) on.

\end{document}